\documentclass[10pt,preprint]{aastex}

\usepackage{graphics}
\usepackage{graphicx}
\usepackage{epsfig}
\usepackage{longtable}
\usepackage{blindtext}
\usepackage{float}

\usepackage{color}
\usepackage{amssymb}
\definecolor{red}{rgb}{1,0,0}
\definecolor{green}{rgb}{0,0.8,0}
\definecolor{blue}{rgb}{0,0,1}

\newcommand\bla{\color{black}}

\shorttitle{Stellar occultation by the TNO 2003~AZ$_{84}$}
\shortauthors{Dias-Oliveira et al.}

\begin{document}


\title{%
Study of the plutino object (208996)~2003~AZ$_{84}$ from stellar occultations:
size, shape and topographic features
}%


\author{%
Dias-Oliveira, A.\altaffilmark{1}, Sicardy, B.\altaffilmark{2}, Ortiz, J. L.\altaffilmark{3}, Braga-Ribas,~F.\altaffilmark{4,1},
Leiva, R.\altaffilmark{2,5}, Vieira-Martins, R.\altaffilmark{1,6,7}, Benedetti-Rossi, G.\altaffilmark{1}, 
Camargo, J. I. B.\altaffilmark{1,7}, Assafin, M.\altaffilmark{8}, Gomes-J\'unior, 
A. R.\altaffilmark{8}, Baug, T.\altaffilmark{9}, Chandrasekhar, T.\altaffilmark{9}, Desmars, J.\altaffilmark{2} 
Duffard, R.\altaffilmark{3}, Santos-Sanz, P.\altaffilmark{3},
Ergang, Z.\altaffilmark{10}, Ganesh, S.\altaffilmark{9}, Ikari, Y.\altaffilmark{11}, Irawati, P.\altaffilmark{12}, Jain, J.\altaffilmark{9},
Liying, Z.\altaffilmark{10}, Richichi, A.\altaffilmark{12}, Shengbang, Q.\altaffilmark{10}, Behrend, R.\altaffilmark{13}, Benkhaldoun, 
Z.\altaffilmark{14}, Brosch, N.\altaffilmark{15}, Daassou, A.\altaffilmark{14}, Frappa, E.\altaffilmark{16}, Gal-Yam, A.\altaffilmark{17},
Garcia-Lozano, R.\altaffilmark{18}, Gillon, M.\altaffilmark{19}, Jehin, E.\altaffilmark{19}, Kaspi, S.\altaffilmark{15}, 
Klotz, A.\altaffilmark{20},
Lecacheux, J.\altaffilmark{2} , Mahasena, P.\altaffilmark{21}, Manfroid, J.\altaffilmark{19}, Manulis, I.\altaffilmark{17},
 Maury, A.\altaffilmark{22},
Mohan, V.\altaffilmark{23}, Morales, N.\altaffilmark{3}, Ofek, E.\altaffilmark{17}, Rinner, C.\altaffilmark{24}, 
Sharma, A.\altaffilmark{25}, Sposetti, S.\altaffilmark{13}, Tanga, P.\altaffilmark{26}, Thirouin, A.\altaffilmark{27}, Vachier, F.\altaffilmark{6}, 
Widemann, T.\altaffilmark{2}, Asai, A.\altaffilmark{11}, Watanabe, Hayato\altaffilmark{11}, Watanabe, Hiroyuki\altaffilmark{11}, 
Owada, M.\altaffilmark{11}, Yamamura, H.\altaffilmark{11}, Hayamizu, T.\altaffilmark{11}, 
Bradshaw, J.\altaffilmark{28,29},  Kerr, S.\altaffilmark{30}, Tomioka, H.\altaffilmark{31},  Andersson, S.\altaffilmark{32}, 
 Dangl, G.\altaffilmark{33},
Haymes, T. \altaffilmark{34}, Naves, R. \altaffilmark{35},  Wortmann, G.\altaffilmark{35}
}%

\email{alexoliveira@on.br}
\email{oliveira.astro@gmail.com}

\altaffiltext{1}{Observat\'orio Nacional/MCTI, Rua General Jos\'e Cristino 77, Rio de Janeiro - RJ, 20.921-400, Brazil}
\altaffiltext{2}{LESIA/Observatoire de Paris, CNRS UMR 8109, Universit\'e Pierre et Marie Curie, Universit\'e Paris-Diderot, 5 place Jules Janssen, F-92195 Meudon C\'edex, France.}
\altaffiltext{3}{Instituto de Astrof\'isica de Andaluc\'ia-CSIC, Apt 3004, 18080, Granada, Spain} 
\altaffiltext{4}{Federal University of Technology - Paran\'a (UTFPR / DAFIS), R. Sete de Setembro 3165, Curitiba - PR, 80230-901, Brazil}
\altaffiltext{5}{Instituto de Astrof\'isica, Facultad de F\'isica, Pontificia Universidad Cat\'olica de Chile, Av. Vicu\~na Mackenna 4860, 7820436 Macul, Santiago, Chile}
\altaffiltext{6}{IMCCE/Observatorie de Paris, 77 Avenue Denfert Rochereau, Paris, 75014, France}
\altaffiltext{7}{Laborat\'orio Interinstitucional de e-Astronomia - LIneA, Rua General Jos\'e Cristino 77, Rio de Janeiro - RJ, 20.921-400, Brazil}
\altaffiltext{8}{Observat\'orio do Valongo/UFRJ, Ladeira Pedro Antonio 43, Rio de Janeiro - RJ, 20080-090, Brazil}
\altaffiltext{9}{Physical Research Laboratory, Ahmedabad, Gujarat 380009, India}
\altaffiltext{10}{Yunnan Observatories,  Chinese Academy of Sciences, Yunnan, China}
\altaffiltext{11}{Japan Occultation Information Network (JOIN), Japan}
\altaffiltext{12}{National Astronomical Research Institute of Thailand, Siriphanich Building, Chiang Mai 50200 - Thailand}
\altaffiltext{13}{Observatoire de Gen\`eve, 1290 Versoix, Switzerland}
\altaffiltext{14}{Oukaimeden Observatory, LPHEA, Cadi Ayyad University, Marroch}
\altaffiltext{15}{Wise Observatory and School of Physics and Astronomy, Tel-Aviv University, Tel-Aviv 69978, Israel}
\altaffiltext{16}{EURASTER, 8 route de Soulomes, 46240 Labastide-Murat, France}
\altaffiltext{17}{Department of Particle Physics and Astrophysics, Weizmann Institute of Science, Rehovot 7610001, Israel}
\altaffiltext{18}{Observatorio de Busot, Spain}
\altaffiltext{19}{Institut d'Astrophysique, de G\'eophysique et Oc\'eanographie, Universit\'e de Li\`ege, All\'ee du 6 ao\^ut 17, 4000 Li\'ege, Belgium }
\altaffiltext{20}{Universit\'e de Toulouse, UPS-OMP, IRAP, 14 Avenue Edouard Belin, 31400 Toulouse, France}
\altaffiltext{21}{Observatorium Bosscha, Institut Teknologi Bandung, Indonesia}
\altaffiltext{22}{San Pedro de Atacama Celestial Explorations (S.P.A.C.E.), San Pedro de Atacama, Chile}
\altaffiltext{23}{IUCAA Girawali Observatory, India}
\altaffiltext{24}{Moroccan Oukaimeden Sky Survey, Marroch}
\altaffiltext{25}{Nikaya Observatory, Tamil Nadu 635114, India}
\altaffiltext{26}{Laboratoire Lagrange,  UMR7293, Uni. C\^ote d'Azur, CNRS, Observatoire de la C\^ote d'Azur Boulevard de l'Observatoire, CS 34229, 06304 Nice, France}
\altaffiltext{27}{Lowell Observatory, 1400 W Mars Hill Rd, Flagstaff, Arizona, United States of America}
\altaffiltext{28}{International Occultation Timing Association (IOTA), PO Box 7152, Kent, WA 98042, USA}
\altaffiltext{29}{Samford Valley Observatory, QLD, Australia}
\altaffiltext{30}{Royal Astronomical Society of New Zealand - Ocultation Section}
\altaffiltext{31}{Independent Astronomer}
\altaffiltext{32}{Amateursternwarte Mueggelheim, Wiesbacher Weg 8 , 12559 Berlin, Germany}
\altaffiltext{33}{A-3830, Nonndorf 12, Austria}
\altaffiltext{34}{BAA Asteroids and Remote Planets Section, Hill Rise, Knowl Hill Common, RG10 9YD, UK}
\altaffiltext{35}{c/Jaume Balmes nº 24 Cabrils 08348, Spain}
\altaffiltext{36}{Archenhold-Observatory, Neue Krugallee 180, D-12437 Berlin, Germany}






\begin{abstract}

We present results derived from four stellar occultations by the plutino object (208996) 2003~AZ$_{84}$,
detected at
January 8, 2011 (single-chord event), 
February 3, 2012 (multi-chord), 
December 2, 2013 (single-chord) and 
November 15, 2014 (multi-chord).
Our observations rule out an oblate spheroid solution for 2003~AZ$_{84}$'s shape.
Instead, assuming hydrostatic equilibrium, we find that a Jacobi triaxial solution with semi axes 
$(470 \pm 20) \times (383 \pm 10) \times (245 \pm 8)$~km
can better account for all our occultation observations. 
Combining these dimensions with the rotation period of the body (6.75~h) and 
the amplitude of its rotation light curve, we derive a density 
$\rho=0.87 \pm 0.01$~g~cm$^{-3}$ a 
geometric albedo $p_V= 0.097 \pm 0.009$.
A grazing chord observed during the 2014 occultation reveals a topographic feature along 2003~AZ$_{84}$'s limb, 
that can be interpreted as an abrupt chasm of width $\sim 23$~km and depth $> 8$~km
or a smooth depression of width $\sim 80$~km and depth $\sim 13$~km
(or an intermediate feature between those two extremes).
\end{abstract}





\section{INTRODUCTION}

The size, mass, shape, albedo, chemical composition and orbital elements distribution of the Trans-Neptunian Objects  (TNOs) 
provide important information on the chemical and dynamical evolution of our Solar System \citep{lykawka08}. 
In addition, as their surfaces are little affected by solar radiation due to their large heliocentric distances, 
they give us clues to the composition of the primordial nebula. 
Unfortunately, the faintness and small angular sizes of most of them, make their study difficult, 
so that our knowledge of these objects remains fragmentary \citep{Stansberry08}.

One of the most effective techniques to study these bodies is the stellar occultation method, 
which can provide sizes and shapes down to km-level accuracy, 
reveal atmospheres at the few nbar level (\citet{Sicardy11, Ortiz12}), 
and detect features like jets, satellites and rings 
\citep{Braga14a,ort15,rup15}.

Besides Pluto and Charon, since 2009, after the first successful observation of a stellar occultation
by the TNO 2002 TX$_{300}$, see \cite{Elliot10}, several
objects have been measured by stellar occultations, like Varuna \citep{Sicardy10}, Eris \citep{Sicardy11},
Quaoar (\citealp{Person11}, \citealp{Sallum11},
\citealp{Braga13}), Makemake \citep{Ortiz12},
2002 KX$_{14}$ \citep{Candal14} and 2007 UK$_{126}$ \citep{Ben16}.

Adding to the list above, we now have (208996) 2003~AZ$_{84}$, a large TNO classified as a plutino (object in the 3:2 mean motion resonance with Neptune) \citep{Marsden09}. 
It was discovered in January 2003 by \citet{Trujillo03} and has an estimated
area-equivalent radius and visible geometric albedo of 
$364_{-33}^{+31}$~km and 0.107 $_{-0.016}^{+0.023}$, 
respectively,  as derived from thermal measurements \citep{mommert12}.  
Its orbit has 
a semi-major axis of 39.406 AU, 
aphelion distance 46.477 AU, 
orbital period 247.37 yr, 
orbital eccentricity 0.179 and 
inclination of 13.563 degrees%
\footnote{http://ssd.jpl.nasa.gov/sbdb.cgi?sstr=2003AZ84}.
It also has a satellite fainter than the primary by $5.0 \pm 0.3$~mag  \citep{Brown07} 
located at a typical distance of 10,000~km from the primary.
However, due to lack of data, its orbit is still unknown and
no mass estimate for the system has been derived.

We present here four stellar occultations by 2003~AZ$_{84}$.
The first one ever observed involving this body was a single-chord event detected on January 8, 2011 in Chile, 
while another single-chord occultation was observed on December 2, 2013 in Australia.
Both events provide lower limits for the semi-major axis of the object, as discussed ahead in the main text.

On the other hand, multi-chord events were observed on February 3, 2012 and on November 15, 2014.
The 2012 event was detected from three sites, two in India and one in Israel.  
The November 15, 2014 event was recorded from  sites in China, Thailand and Japan.
The latter provides the best signal-to-noise ratio and time resolution light curves of all four events.
Moreover, the China observation shows a gradual star disappearance that is better fitted
as a topographic feature at the surface of the body.
This is the first solid report of of this kind from a stellar occultation by a TNO.

The occultation prediction methods are outlined in Section~\ref{sec:PREDICTIONS-AND-OBSERVATIONS}.
Data analysis is  presented in Section~\ref{sec:DATA-ANALYSIS} and 
results are given in Section~\ref{sec:RESULTS} 
(in particular on the size and shape of the body and on the presence of a topographic feature), 
before concluding remarks in Section~\ref{sec:CONCLUSIONS}.

\section{PREDICTIONS AND OBSERVATIONS\label{sec:PREDICTIONS-AND-OBSERVATIONS}}


Considering the importance of TNO occultations and the challenges they pose
(essentially due to the high degree of accuracy required on the TNO orbit and star position), 
huge efforts were dedicated to the prediction of the events described here.
The candidate stars were identified in systematic surveys performed at the 2.2~m telescope of 
the European Southern Observatory (ESO) at La Silla, using the Wide Field Imager (WFI). 
The surveys yielded local astrometric catalogues for 5 Centaurs and 34 TNOs (plus Pluto and its moons) up to 2015, 
with stars with magnitudes as faint as $R_{\rm mag} \sim 19$ mag \citep{Assafin10,Assafin12,Camargo14}. 
Close to the occultation dates, astrometric updates of the candidate stars (and if possible, the TNO)
were performed to improve the predictions by avoiding systematic biases, like catalog errors.


The updates for the February 3, 2012 event were based on observations carried out between September 22, 2011 and January 25, 2012 
(Table~\ref{tab_astrometry_2012}). 
Those measurements confirmed that the shadow's path should pass across North Africa and Middle East.
They provide the following ICRF/J2000 star position:
\begin{eqnarray}
\alpha=07^{\rm h} 45^{\rm m} 54.7696^{\rm s}\pm0.022'' \\
\delta=+11^{\circ}12'43.093''\pm0.023'', \nonumber 
\label{eq:a_d_star_2012}
\end{eqnarray}

\begin{figure*}[!htb]
\centering
\includegraphics[height=40mm,trim= 00 0 00 120]{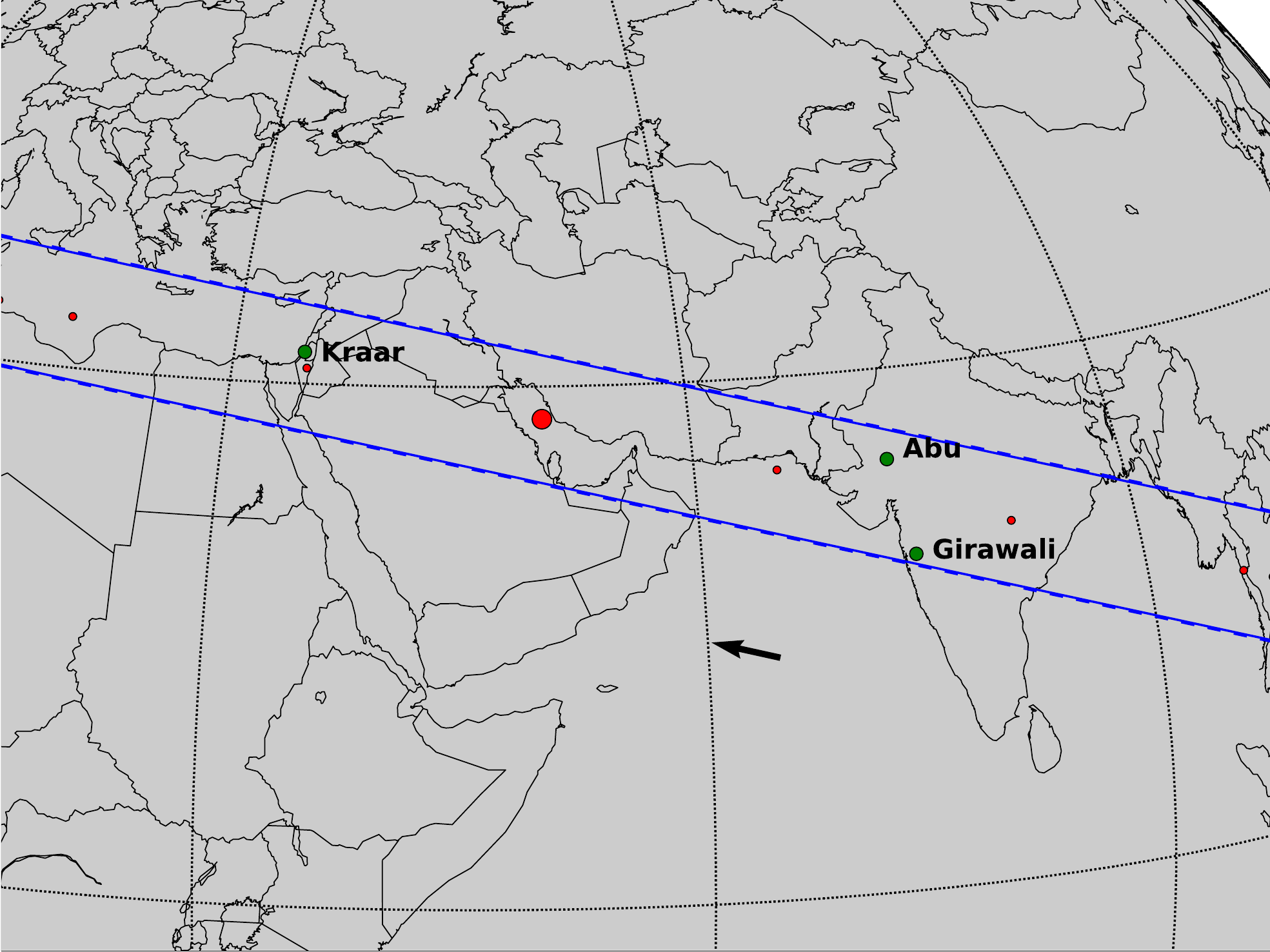}
\includegraphics[height=65mm,trim= 30 150 -100 160]{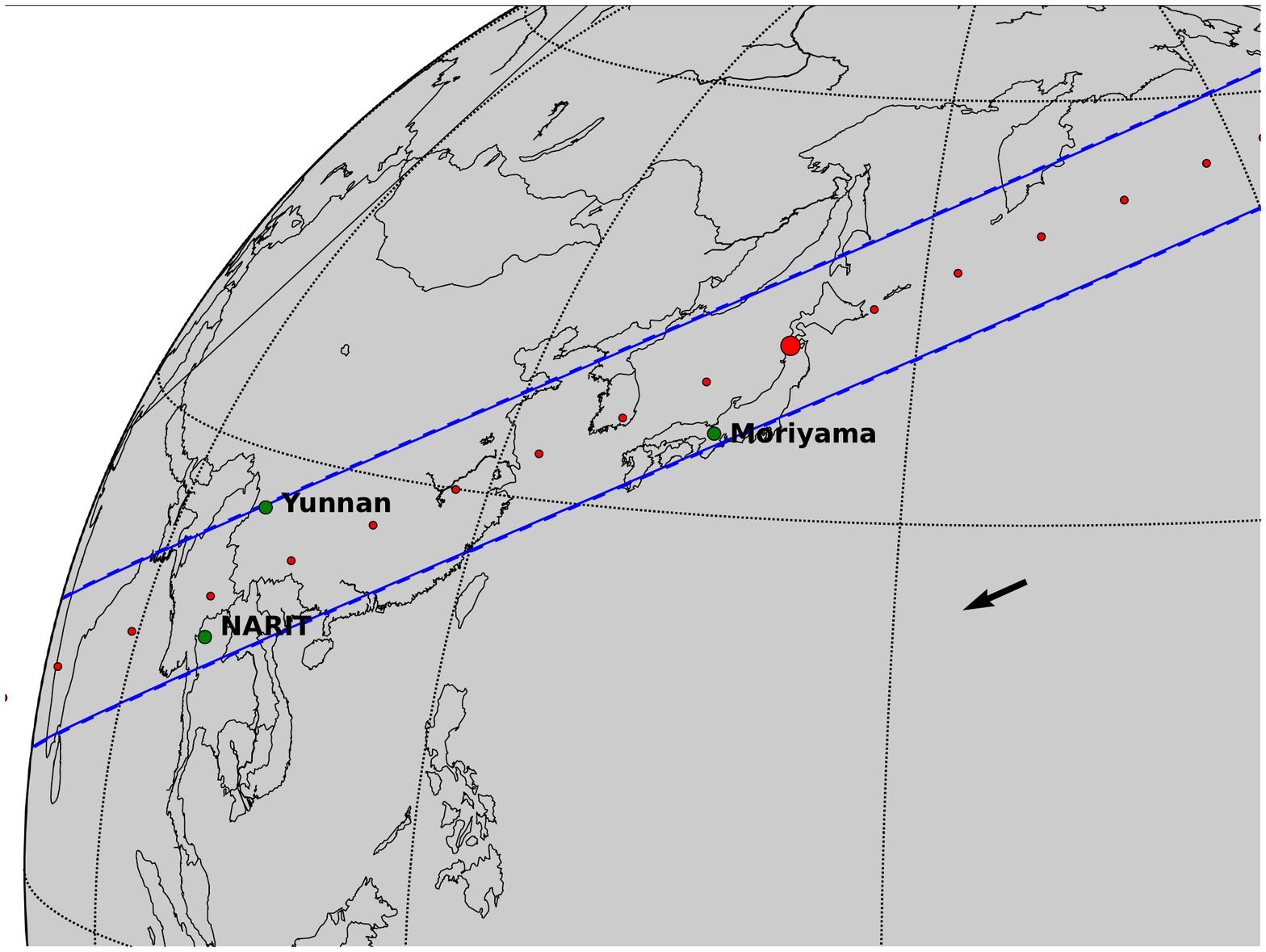}
\caption{%
\normalsize
The shadow paths of the two multi-chord events analyzed here
(February 3, 2012 at left and November 15, 2014 at right), 
the blue solid lines corresponding to the approximate size of the object. 
The smaller red dots mark the shadow center every minute, 
while the larger dot indicates the time of closest approach to the geocenter. 
The arrows show the direction of motion.
The green dots mark the sites where data were obtained with positive detection of the event.
}%
\label{fig:PredicMap_2012_2014}
\end{figure*}

\textbf{For the November 15, 2014 event, we used the Version 1 of NIMA ephemeris%
\footnote{Numerical Integration of the Motion of an Asteroid, 
see http://josselin.desmars.free.fr/tno/2003AZ84}
 (NIMA V1 hereafter) to get a short period correction for
2003~AZ$_{84}$'s motion, using previous observations. NIMA determines and propagates the orbit through the numerical integration of the
equations of motion, based on the least squares method to iteratively find the corrections to each component of the state
vector from observations. Compared to JPL ephemeris, we used additional observations from Pic du Midi and ESO, two astrometric
positions deduced from positive occultations on 8 January 2011 and 3, February 2012, and a specific weighting scheme depending on the
observatory, the stellar catalogue used for the reduction, and the number of observations per night (see \citealt{Demars15} for more
details). } 
The ephemeris, with average standard deviations of 40 milli-arcsec (mas) with respect to observations, 
was  combined with astrometric observations of the candidate star, 
carried out with the two 0.60~m telescopes (Zeiss and Boller and Chivens) 
at Pico dos Dias Observatory in Brazil (IAU code 874), providing the following ICRF/J2000 star position:
\begin{eqnarray}
\alpha= 08^{\rm h}03^{\rm m}51.2980^{\rm s}\pm0.022'' \\
\delta=+09^{\circ}57'18.729''\pm0.023''\nonumber
\label{eq:a_d_star_2014} 
\end{eqnarray}
The combination confirmed a shadow path going over Japan, Thailand and part of China, 
leading us to trigger alerts at several potential sites shown in Fig.~\ref{fig:PredicMap_2012_2014}.
The circumstances of observations for both 2012 and 2014 events are listed in 
Tables~\ref{tab_circum_2012} and~\ref{tab_circum_2014}.

\section{DATA ANALYSIS\label{sec:DATA-ANALYSIS}}

\subsection{OCCULTATION LIGHT CURVES
\label{sub:OCCULTATION-LIGHT-CURVES}}

For the events described below,
the flux from the star (plus the faint, background contribution due to the occulting body)
was obtained through differential aperture photometry using the PRAIA package \citep{Assafin11}.
The resulting light-curve (flux vs. time) was then normalized to its unocculted value,
using polynomial fits -- first or third degree depending on the data quality -- to the flux before and after the event.

\subsubsection{Single-Chord events 2011 and 2013}

\begin{center}
\begin{figure}[!htb]
\centering
\includegraphics[height=70mm]{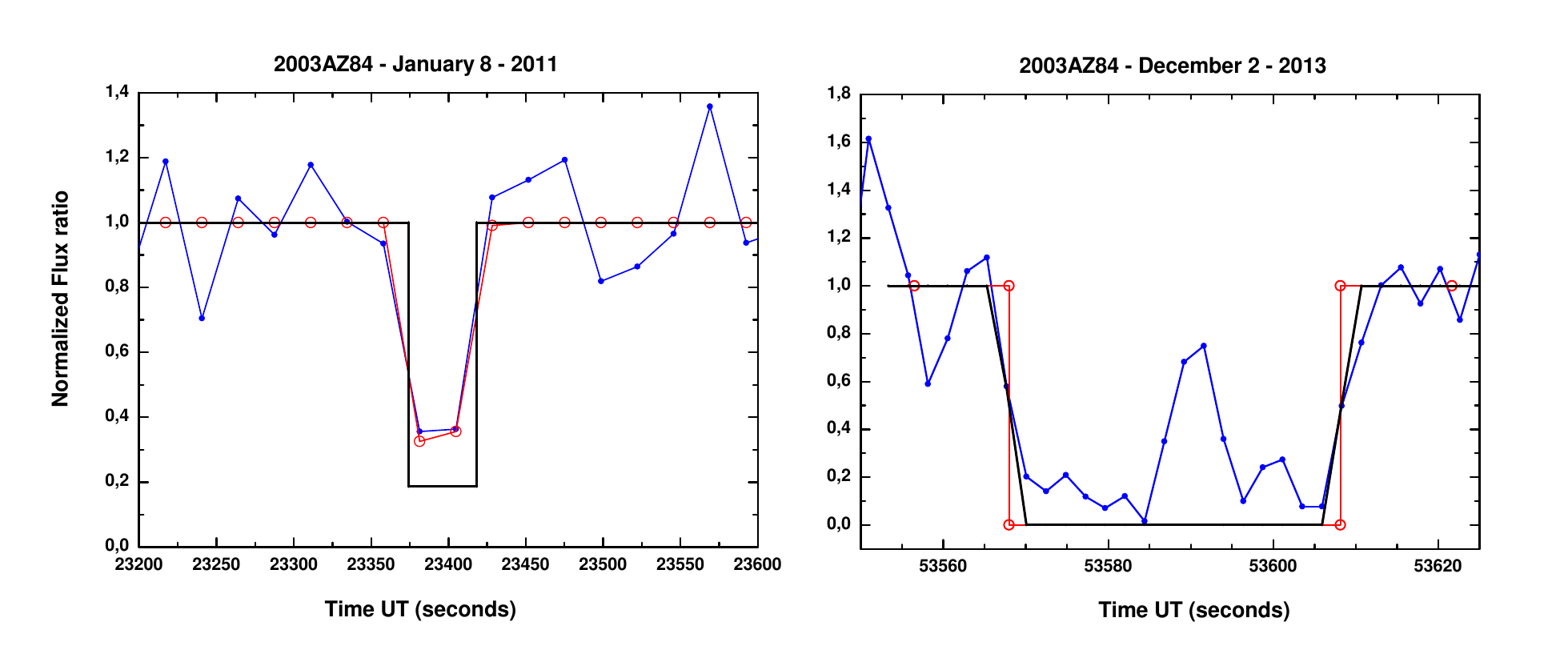}
\caption{%
Fits to the single-chord events of January 8, 2011 (left panel) and December 2, 2013 events (right panel).
In both panels, the blue lines and dots are the observed occultation light curve plotted vs. time.
The solid black lines show the square-well models that best fit the data, while
the red lines and circles show the synthetic flux after convolution of the square-well models by
Fresnel diffraction, star angular diameter projected at the body and finite exposure time,
see Section~\ref{sub:OCCULTATION-TIMING} for details.
}%
\label{fig:lightcurves_2011_2013}
\end{figure}
\end{center}

The  single chord occultation of January 8, 2011 had a typical shadow velocity of 26~km~s$^{-1}$.
It was observed at San Pedro de Atacama Celestial Explorations Observatory 
(longitude $68^{\circ}$ 10' 48.70'' W, latitude $22^{\circ}$ 57' 09.80'' S, altitude 2400 m)
in Chile, by Alain Maury using the C. Harlinten 0.5-m Planewave telescope (Fig.~\ref{fig:lightcurves_2011_2013}) and by Nicolas Morales
with the remotely operated 0.4-m ASH2 telescope.
On the other hand, the single chord event of December 2, 2013 had a shadow velocity of about 17~km~s$^{-1}$ and 
was observed by Steve Kerr,  near Rockhampton
(longitude $150^{\circ}$ 30' 00.80'' E, latitude $23^{\circ}$ 16' 09.60'' S, altitude 50 m)
in Australia. 
The event was recorded with a Watec 120N+ video camera attached to a 0.30~m telescope attached, 
with cycle 2.56~s exposure time. 

The occultation durations (see Fig. \ref{fig:lightcurves_2011_2013} and Table~\ref{tab_ingress_egress}) 
corresponds to chord lengths of 
$573 \pm 21$~km and $657 \pm 35$~km for the 2011 and 2013 events, respectively 
and provide lower limits for the major-axis of the object \citep{Braga11,Braga14b}.

\subsubsection{February 3, 2012}

Three occulting sites recorded this event, Mount Abu Observatory and IUCAA (Inter-University Centre for Astronomy and Astrophysics) Girawali 
Observatory in India, plus the Kraar Observatory in Israel. The event had a typical shadow velocity of 25~km~s$^{-1}$,
see details in Table~\ref{tab_circum_2012}.

\textbf{Although all the sites had robust clock synchronizations, 
the acquisition software used for all of them recorded only the integer part of the second in each image header. }
\textbf{In order to retrieve the fractional part of the second for the mid-exposure time $t_i$ of each image $i$, 
we performed linear fits to the set $(i, t_i)$ as used by \cite{Sicardy11}.}

\begin{center}
\begin{figure}[!htb]
\centering
\includegraphics[height=70mm]{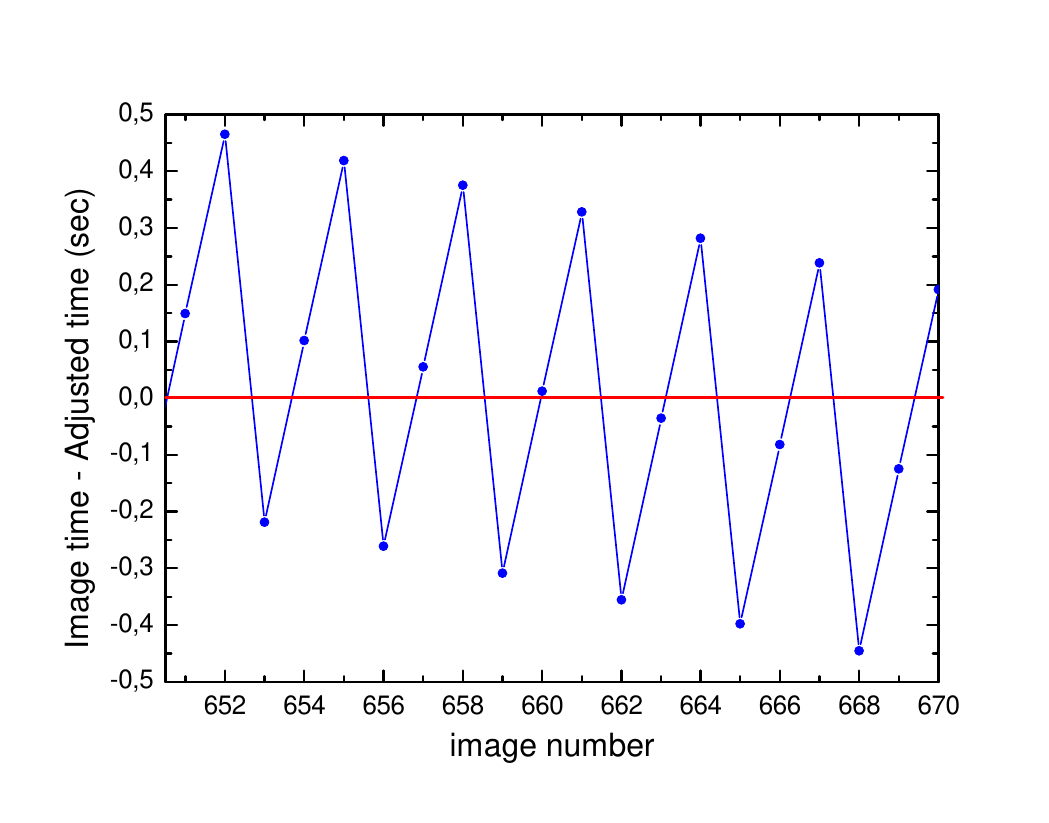}
\caption{%
Example of the saw pattern obtained with the residuals of the linear fit (blue line) of the truncated time.  
For this kind of fit we used points around the event, during a period in which the acquisition was regular.
}%
\label{time_fit}
\end{figure}
\end{center}

\textbf{The residuals of this linear fit show a saw pattern with dispersion between -0.5 and +0.5 seconds (Fig. ~\ref{time_fit} for example),
 showing that there is 
truncation and the period in which the acquisition cycle is regular. From this fit we obtain the acquisition cycle $t_c$, which
 is used to obtain the time of each image $t_{fit}$, now with the fraction of the second, and an initial time $t_0$ (Eq. ~\ref{eq:time_fit}).} 

\begin{eqnarray}
t_{fit} = t_0 + (t_c \cdot i) + 0.5
\label{eq:time_fit} 
\end{eqnarray}

\textbf{Half a second is added to the times to correct the offset imposed by the truncation. With this procedure the times of
 each image have an internal precision that depends on the square root of the number of images used
and is less than one second.  In the present case, we could retrieve
the individual times $t_{fit}$ to within an accuracy of  0.06~s.}

Note that the IUCAA light curve was obtained with an exposure time of 2~s, and a long read-out time of 14.5~s
which provides a single occultation point with partial flux drop, and a large uncertainty on the ingress time at that station,
see discussion below, Table~\ref{tab_ingress_egress} and Fig.~\ref{fig:elipse_fit_2012}.

\subsubsection{November 15, 2014}

\begin{center}
\begin{figure}[!htb]
\centering
\includegraphics[scale=1.1]{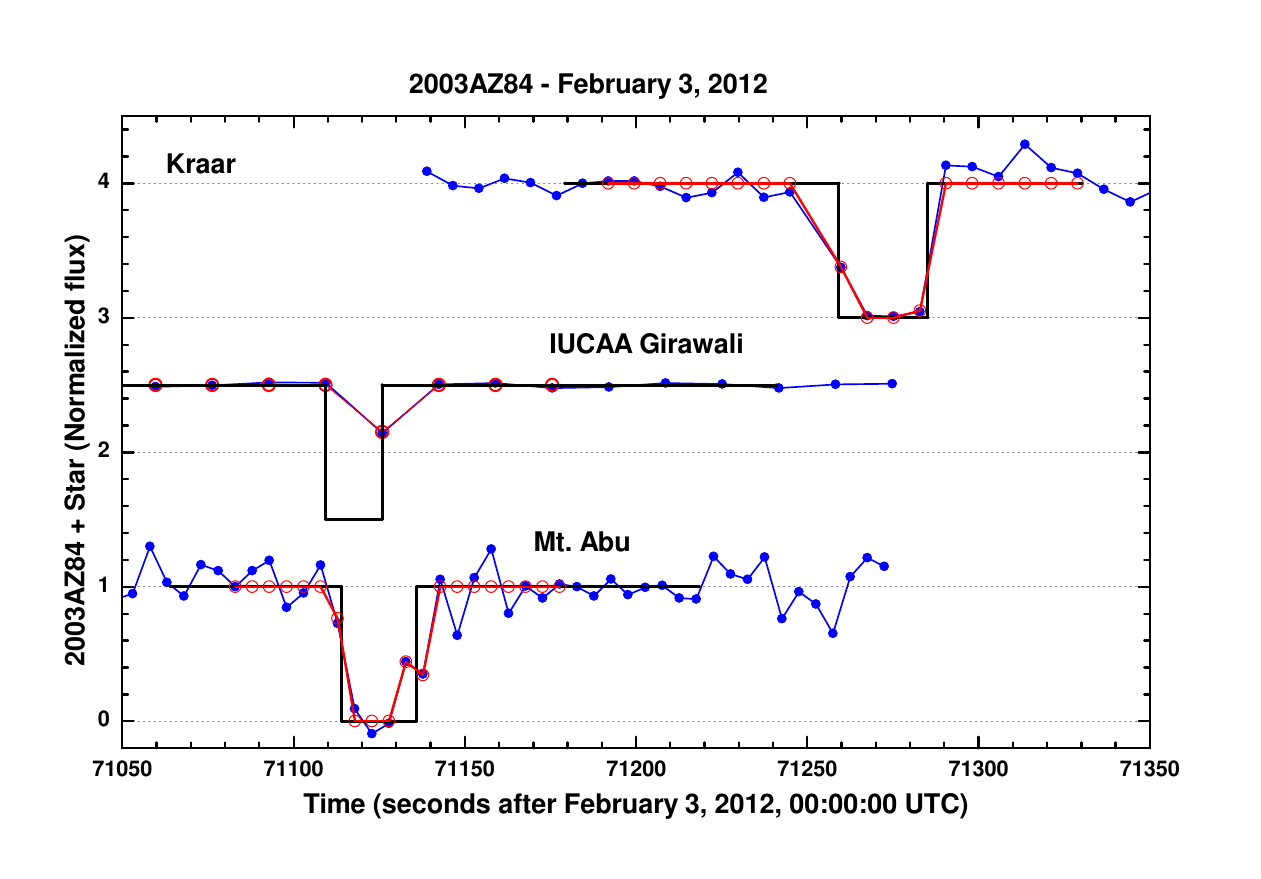}
\includegraphics[scale=1.1]{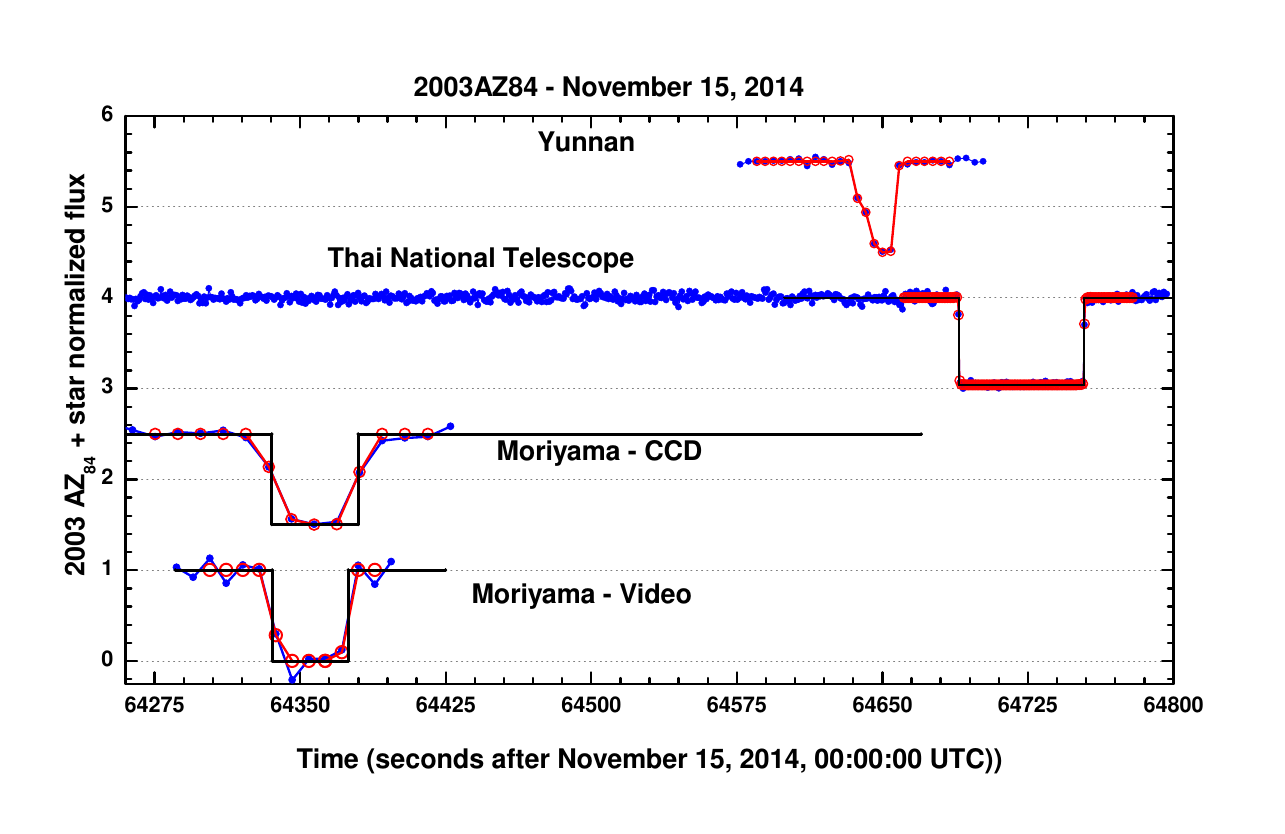}
\caption{%
Same conventions as for the  previous figure.
Top panel: the four occultation light curves obtained during the 2014 event,
vertically shifted for better viewing. 
Bottom panel: same as top panel for the three light curves obtained during the 2012 event.
}%
\label{fig:lightcurves}
\end{figure}
\end{center}

This event was the slowest of all described here, with a shadow velocity of about 9.4~km~s$^{-1}$.
Four light curves were obtained at three different sites:
the Yunnan station in China, 
the Thai National Telescope (TNT) in Thailand and 
the Moriyama station in Japan,
see Table~\ref{tab_circum_2014} and Fig.~\ref{fig:lightcurves}. 
For the three CCD recordings, the mid-exposure times $t_i$ were extracted from the image headers. 
The headers from the Thai National Telescope images provide $t_i$ down to the millisecond 
while for both the Yunnan and Moriyama images, only the integer part of the second was recorded,
thus requiring the same procedure as described in the previous subsection.

The Moriyama site also recorded the event with a video camera attached to a 25.4~cm telescope.
The data were collected by integrating 256 video frames to provide a complete exposure of approximately 8~s, 
with GPS time inserted in each frame. 
Using the Audela software\footnote{%
a free and open source astronomy software for digital observations: CCD cameras,
Web cams, etc..., see http://audela.org/dokuwiki/doku.php?id=en:start},
we extracted individual frames in FITS format from the video. 
This conversion process requires special care because of possible dropped frames, 
duplicated fields or any software incompatibility \citep{Buie16}. Because of that, detailed checks were performed in all sets of images to ensure that the extracted
time corresponds to the time printed at each frame.

Also, AUDELA conversion software create a FITS image for each video frame, instead of for each exposure. 
So, it is important to check exactly how many of the exported FITS frames belong to a single acquisition, 
in order to combine them properly into a single image.
Using photometry and observing the step-like variations of the target brightness, 
we can identify in each step the 256 frames that represent a single acquisition and that must be averaged out.

More precisely, in order to avoid taking into account any field from the previous or next acquisition, 
we excluded the first and last frame from a sequence, and then averaged out the remaining 254 frames. 
To ensure a correct time-stamping of each final images, we used the mid-exposure time of the 
mid-frame of each sequence.


A noteworthy feature observed during this occultation is the gradual decrease of the stellar flux observed over three 
acquisition intervals during the star disappearance at the Yunnan site (Fig. \ref{fig:yunnan_fit}). 
This feature is analyzed in more detail below and was interpreted as a topographic feature.
%
%
Note that this effect cannot be the result of photometric noise, as
the typical photometric error of each data point is about 0.0236 mag, less than 6\% of the observed flux variation, 
thus ruling out photometric problems. 
Moreover, a careful visual analysis of the three corresponding images clearly shows a gradual disappearance of the star,
without any similar behavior for the other reference objects acquired in the field of view. 

\subsection{OCCULTATION TIMING
\label{sub:OCCULTATION-TIMING}}

To determine the start and end times of the occultation we fitted for each light curve a sharp edge occultation model 
convolved by 
Fresnel diffraction, 
CCD bandwidth, 
stellar diameter projected at the body and 
finite integration time, see \citet{Widemann09, Braga13}.

With 2003~AZ$_{84}$'s geocentric distances of 
$D = 44.3$~AU  in 2012 and  
$D = 44.5$~AU  in 2014,
the Fresnel scale ($F=\sqrt{\lambda D/2}$) for a typical wavelength of $\lambda=0.65$ $\mu$m (and $\lambda=1.65$ $\mu$m for Mt. Abu)
is derived for each event, 
$F =1.47$~km (2.34~km for Mt. Abu) in 2012, and 
$F=1.46$~km in 2014.

The star diameter projected at 2003~AZ$_{84}$'s distance was estimated using 
the $B$, $V$ and $K$ apparent magnitudes provided by the NOMAD catalogue \citep{zacharias04} and
the formulae of \citet{van99}.
The 2012 star ($B=15.8$ mag, $V=15.35$ mag, $K=14.1$ mag) has then an estimated projected diameter of 0.24 km, 
while it is 0.20 km for the 2014 star ($B=15.5$ mag, $V=15.5$ mag and $K=14.0$ mag).

Using the apparent TNO motion relative to the star, 
it is possible to translate the integration times into actual distances traveled in the sky plane. 
The smallest integration times were 2.0~s in 2012 and 0.786~s in 2014,
corresponding to 49.7~km and 7.50~km in the celestial plane, respectively. 
Therefore, our light curves are dominated by the integration times, and not by Fresnel diffraction or stellar diameter. 
The same is true for the 2011 and 2013 events.

The free parameters to adjust in the fits are the times $t_{\rm occ}$ of ingress (star disappearance) or egress 
(star re-appearance) for each station. 
The value of $t_{\rm occ}$ is obtained by minimizing a classical $\chi^{2}$ function, as described in \citet{Sicardy11}. 
The resulting best fits are shown in Fig.~\ref{fig:elipse_fit_2012}, 
and the occultation times are listed in Table \ref{tab_ingress_egress}.

The IUCAA light curve is special, however, because it has only one data point with a partial drop of the stellar flux 
(Fig.~\ref{fig:elipse_fit_2012}), meaning that the star was occulted during a fraction of this 2-s acquisition interval.
Calibration data taken with the same instrument on February 10, 2012 provide the separate fluxes of 2003~AZ$_{84}$ and the star, 
allowing us to estimate the fraction of time when the star was occulted during that interval.
Note that from the light curve, we cannot discriminate between an ingress or egress point.
However, considering the location of the point in the sky plane and comparing it to the other chords, 
it seems more likely to be an egress point, with an ingress that happened during 14.5 s gap between
the end of the previous exposure, and before the start of the current exposure. 
The alternative  solution (with egress happening during the gap after the single drop)
would result in a very elongated body which is unlikely, see Fig.~\ref{fig:elipse_fit_2012}.

\subsection{LIMB FITTING}
\label{sub:LIMB-FITTING}
 
The ingress and egress times $t_{\rm occ}$
provide the positions of the star in the sky plane relative to 2003~AZ$_{84}$'s center, 
once the star position is specified (Eqs.~\ref{eq:a_d_star_2012} and \ref{eq:a_d_star_2014}),
and for a given ephemeris (NIMA~V1 in our case).
More precisely, each timing gives the position $(f,g)$ of the star with respect to the body center,
$f$ and $g$ being measured positively towards local celestial east and celestial north, respectively.

For each event, the general shape for the body's limb was assumed to be an ellipse 
characterized by $M=5$ adjustable parameters:
the coordinates of the ellipse center, $(f_c,g_c)$; 
the apparent semi-major axis $a'$; 
the apparent oblateness $\epsilon'=(a'-b')/a'$ (where $b'$ is the apparent semi-minor axis) and 
\textbf{the position angle of the pole $P_p$ of $b'$, which is the geocentric position
angle of the pole measured eastward from the north.}
Note that the center $(f_c,g_c)$ actually measures the offsets in right ascension and declination 
to be applied to the ephemeris in use, assuming that the star position is correct.
Note also that the quantities $a'$, $b'$, $f_c$ and $g_c$ are all expressed in kilometers.

The ingress and egress times from each light curve provides the extremities of the corresponding occultation chord. 
As both ingress and egress times from Moriyama's light curves agree at 1-$\sigma$ level, 
we consider only one chord for that station (the one obtained with CCD) for the limb fitting.
Thus, for each event there are three positive observations, providing $N=6$ chord extremities (ingress and egress), 
with positions $f_{i,obs}$, $g_{i,obs}$ (Fig.~\ref{fig:elipse_fit_2012} and ~\ref{fig:elipse_fit_2014}).  
The best elliptical fit to the chord extremities is found by minimizing the relevant $\chi^{2}$ function
(see e.g. \citealt{Sicardy11}). 
In that case, the statistical significance of the fit is evaluated from the $\chi^{2}$ per degree of freedom (pdf) 
defined as $\chi_{\rm pdf}^{2}=\chi^{2}/(N-M)$, which should be close to unity. 
\textbf{The individual 1-$\sigma$ error bar of each parameter is obtained by varying manually the given parameter from its nominal solution
value (keeping the other ones fixed) so that $\chi^{2}$ varies from its minimum value  $\chi_{\rm min}^{2}$ to $\chi_{\rm min}^{2}+1$.}

\section{RESULTS\label{sec:RESULTS}}

\subsection{LIMB FITTINGS TO THE 2012 AND 2014 EVENTS}

The parameters of the best elliptical fits for the 2012 and 2014 events are given in Table~\ref{tab_results},
which provides the shapes of the limb and the offsets to be applied to the used ephemeris.
Those fits clearly show a change in the limb shape between those two dates, 
see Figs.~\ref{fig:elipse_fit_2012} and \ref{fig:elipse_fit_2014}.
This said, the fits provide best values of $\chi_{\rm pdf}^{2}=0.73$ (2012) and $\chi_{\rm pdf}^{2}=0.98$ (2014), 
which indicates satisfactory fits.

The main change concerns to the apparent oblateness of the body that varies from
$\epsilon'=0.340^{+0.097}_{-0.086}$ in 2012 to
$\epsilon'=0.054\pm0.003$ in 2014, with a difference of more than 3-$\sigma$.
Conversely, the apparent semi-major axes are $a=426 \pm 20$~km (2012) and $a=393 \pm 1$ km (2014),
with a barely significant difference at 2-$\sigma$ level.
Note in passing that these results agree with the lower limits on the major-axis $573 \pm 21$~km and $657 \pm 35$~km
quoted earlier, derived from the single-chord events of 2011 and 2013 events.
Similarly, the two fits provide consistent (1-$\sigma$ level) position angles of the limb:
$P_L={41^{\circ}}^{+8}_{-5}$ (2012) and $P_L=36^{\circ}\pm4^{\circ}$ (2014). The $P_L$ is geocentric position angle of 
the northern semi-minor axis of the projected limb, measured eastward from the north. (not to be confounded with 
\textbf{position angle of the pole $P_p$}).

We may attribute the change of the retrieved $\epsilon'$ to a very irregular shape, 
but reconciling the limbs of Figs.~\ref{fig:elipse_fit_2012} and \ref{fig:elipse_fit_2014} at face values
would require topographic features of up to 40~km above an average elliptical limb.
Hydrostatic equilibrium is expected for objects with diameter on the order of 1000 km.
More precisely, the critical diameter necessary to reach equilibrium for bodies with slow rotation may vary 
between 500--1200 km for rocky bodies, and 
between 200--900 km for icy bodies \citep{tancredi08}. 
In this context, the body assumes either the shape of a 
Jacobi ellipsoid with principal axes $a>b>c$ or that of a 
Maclaurin spheroid $a=b>c$, depending on their angular momentum \citep{chandra87}. 

Assuming first a Maclaurin solution, 
the variation of $\epsilon'$ between 2012 and 2014 could in principle be caused by 
a variation with time of the opening angle%
\footnote{The angle between the line of sight and the equatorial plane of the body, 
with $B=0^{\circ}$ (resp. $B=90^{\circ}$) corresponding to the equator-on (resp. pole-on) geometry.
It is sometimes referred to as the observer planetocentric elevation and 
is related to the polar aspect angle $\xi$ through $B= 90^\circ - \xi$.}
$B$ due to a mere change of viewing geometry. 
The true oblateness $\epsilon=1-(c/a)$ is related to the apparent oblateness through
$\epsilon'=1-\sqrt{\sin^{2}(B)+(1-\epsilon)^{2}\cos^{2}(B)}$,
and it is easy to see that the larger $\epsilon$, the smaller the change of $B$ necessary to
change $\epsilon'$ by a given amount. 
As the Maclaurin solutions have $\epsilon < 0.417$ \citep{tancredi08}, 
it results that changing $\epsilon'$ from $0.340^{+0.097}_{-0.086}$ to $0.054\pm0.003$
requires changing of $B$ by at least $45^{+20}_{-15}$~degrees.
This cannot be caused just by a variable viewing geometry, 
because of the large geocentric distance of 2003~AZ$_{84}$ (more than 43~AU)
and of its slow heliocentric motion.
In fact, testing a set of random pole positions, the maximum difference in $B$ between 2012 and 2014 
is found to be less than $\sim$~5~degrees. 

An alternative explanation for a changing $B$ is a precession caused by the solar torque on 2003~AZ$_{84}$.
Classical calculations (e.g. \citealt{gol50}) provide a precession rate of
$-3GM_\odot/(2r^3\omega)\epsilon \cos(\theta)$ for a spheroid, 
where
$G$ is the constant of gravitation, 
$M_\odot$ and $r$ are the Sun mass and distance, respectively, 
$\omega$ is the spin frequency, 
and $\theta$ is the obliquity.
At $r > 43$~AU and with a rotation period of $6.75 \pm 0.04$~h \citep{thirouin10}, we have
$3GM_\odot/(2r^3\omega) \sim 10^{-15}$~rad~s$^{-1}$, 
far too small to explain any significant change of pole orientation in only two years.

The same conclusion follows if the precession is caused by the satellite reported by \cite{Brown07}. 
From the numbers quoted in the Introduction, and 
assuming same albedos, the satellite should have a typical radius of 40~km,
corresponding to a mass $M$ of a few times $10^{17}$~kg for an icy composition.
This implies $3GM/(2r^3\omega) \sim10^{-10}$~rad~s$^{-1}$ (taking $r \sim 10,000$~km), 
still too small for explaining any significant changes of the pole orientation in two years.

Finally, 
the object aspect angle might change due to a misalignment between 2003~AZ$_{84}$'s angular
momentum and its minor axis, causing a wobbling, for instance as a consequence of a collision.
\citet{burns73} estimated alignment times for typical small bodies from 4 to 200 km in radius,
and found typical times from $\sim 6 \times 10^7$ to $\sim 10^5$ yr,  respectively. 
Therefore, for 2003~AZ$_{84}$, the alignment time should be very short
($< 10^5$ years), making the wobbling hypothesis highly unlikely.
At this point, we are left with the Jacobi shape alternative, which is explored in more details in the next section.

%
%

\subsection{JACOBI SOLUTION}

Here, we look for a unique Jacobi body with fixed pole position that
can account for both 2012 and 2014 occultations (Figs.~\ref{fig:elipse_fit_2012} and \ref{fig:elipse_fit_2014}).
A second condition that we impose is that 
this solution is also compatible with the (single peaked) amplitude of 2003~AZ$_{84}$'s rotational light curve, 
$\Delta m= 0.07 \pm 0.01$ mag reported by \cite{thirouin10,ortiz06}.

For a stable Jacobi ellipsoid, 
the adimensional quantity $\Omega=\omega^2/(G\pi\rho)$ is bounded between 0.284 and 0.374 \citep{chandra87}.
If we assume that the body is a triaxial ellipsoid, 
the full rotation period should actually be $6.75 \times 2 =  13.5$~h.
This would imply, from the definition of $\Omega$, that the density lies in the interval
$0.21 < \rho < 0.28$~g~cm$^{-3}$, unrealistic low values for $\rho$.
If we adopt a rotational period of 6.75~h (thus assuming that the rotational light curve is caused by
albedo features, not by shape effects, see discusion below), 
then the density is restricted to the interval $0.85 < \rho < 1.12$~g~cm$^{-3}$.

Now, for each value of $\rho$, there is a unique Jacobi ellipsoid solution, and more precisely unique ratios $b/a$ and $c/b$,
where $a>b>c$ are the principal axes of the body.
We simplify the problem further by assuming that the object does not change its pole orientation relative to the observer 
between the 2012 and 2014 occultations.
In this way, the changes of its limb shape are solely due to the rotation around the pole axis.
As commented before, exploring random pole orientations, 
the maximum difference in $B$ between 2012 and 2014 is less than $\sim$~5~degrees, 
while the typical change in the position angle of the pole is $\sim$~7~degrees. 
making our assumption reasonable.

Then, we determine numerically a set of combinations of $B$ and $\rho$ for which there is a pair of rotation angle%
\footnote{The rotation angle origin, $Q=0$, is arbitrarily defined when the longer axis \textit{a} points towards the observer and 
is counted positively following the right-hand rule around the pole axis.}~$Q$
and common value of $a$ that is compatible (to within error bars) with the projected sizes and oblatenesses $\epsilon'$ 
of both 2012 and 2014 events.
Note that given the current accuracy on 2003~AZ$_{84}$'s rotation period ($6.75 \pm 0.04$~h) and the
interval of more than two years between the two observations, 
the corresponding values of $Q$ are considered as uncorrelated between 2012 and 2014. 

Next, for a set of densities $\rho$ between 0.85 and 1.12~g~cm$^{-3}$, 
we consider the corresponding Jacobi solutions, with their associated ratios $b/a$ and $c/a$.
This constrains $(\rho,B)$ to lie on a banded-shape domain that reflects
the 1-$\sigma$
uncertainties on the limb parameters, see the hatched region in
Fig.~\ref{fig:jacobi_opening_angle_vs_density}.
We note that near equator-on solutions ($B \sim 0^{\circ}$) are ruled out as they generate very elongated projected shapes, 
especially for low density ellipsoids, that are incompatible with the moderate apparent oblatenesses of Table~\ref{tab_results} (and with the
 very small amplitude  derived from the rotational light curve). 
On the other hand, near pole-on solutions ($B \sim 90^{\circ}$) are also ruled out as they generate little changes
in projected oblateness whatever the value of $Q$ is, while the oblatenesses between 2012 and 2014 are significantly different.

We now use the amplitude of the rotational lightcurve ($\Delta m= 0.07 \pm 0.01$ mag) as an independent constraint.
Since the 6.75~h rotational light curve is single peaked, while we assume here a Jacobi shape, this means that the 
observed variability is caused by albedo features, not by the shape. 
In that context, we can only state that the Jacobi solution must be such that it causes variations
of apparent cross section that have an amplitude smaller than 0.07 magnitude \citep{bin89}
This defines another possible domain for $(\rho,B)$,
see the gray band in  Fig. \ref{fig:jacobi_opening_angle_vs_density}.
%
The intersection of the two domains puts a stringent constraint 
$\rho=0.87 \pm 0.01$~g~cm$^{-3}$ for the density of the Jacobi body we are looking for,
with axis ratios $b/a=0.82\pm0.05$ and $c/a=0.52\pm 0.02$.
Exploring the intersection domain shown in Fig.~\ref{fig:jacobi_opening_angle_vs_density},
we find that Jacobi ellipsoids with a fixed pole orientation can fit both 2012 and 2014 limbs 
with a semi axes ranges of 
$a=470 \pm 20$~km, 
$b=383 \pm 10$~km and 
$c=245 \pm 8$~km 
(1-$\sigma$ level).

An example of a possible solution is given in Fig.~\ref{fig:representative_jacobi_solution},
corresponding to $a=456$~km, $\rho=0.86$~g~cm$^{-3}$ and 
rotation angles $Q=24^{\circ}$ (2012) and $Q=10^{\circ}$ (2014). 
A full treatment that provides a global best fit and finer error bars would require a bayesian approach.
This remains out of the scope of this paper, but our example shows that a unique Jacobi solution
meeting all the constraints (our occultation results plus the rotational light curve amplitude) 
do exist in narrow intervals of $\rho$ and $a$.

Finally, the intersection region in Fig.~\ref{fig:jacobi_opening_angle_vs_density}
provides a range for the equivalent radius  of the body, 
$R_{\rm eq} = 386\pm6$~km (as observed in 2012 and 2014),
i.e. the radius of the disk that has the same apparent area as the observed body.
Using this value, we obtain the geometric albedo in visible band through
\begin{equation}
p_V=(AU_{\rm km}/R_{\rm eq})^{2}\times10^{0.4(H_{\odot,V}-H_V)},
\label{eq:albedo}
\end{equation}
where $AU_{\rm km}=1.49597870700\times10^{8}$ km, $H_{\odot,V}=-26.74 $ mag is the Sun magnitude at 1 AU in the visible.
From 2003~AZ$_{84}$'s visible absolute magnitude, $H_{V}=3.74\pm0.09$ mag \citep{mommert12},
and the value $R_{\rm eq}$ obtained above, we derive  $p_{V}= 0.097 \pm 0.009$. 

\subsection{TOPOGRAPHIC FEATURE}
\label{sub:TOPOGRAPHIC}

\begin{center}
\begin{figure}[!htb]
\centering
\includegraphics[scale=1.2]{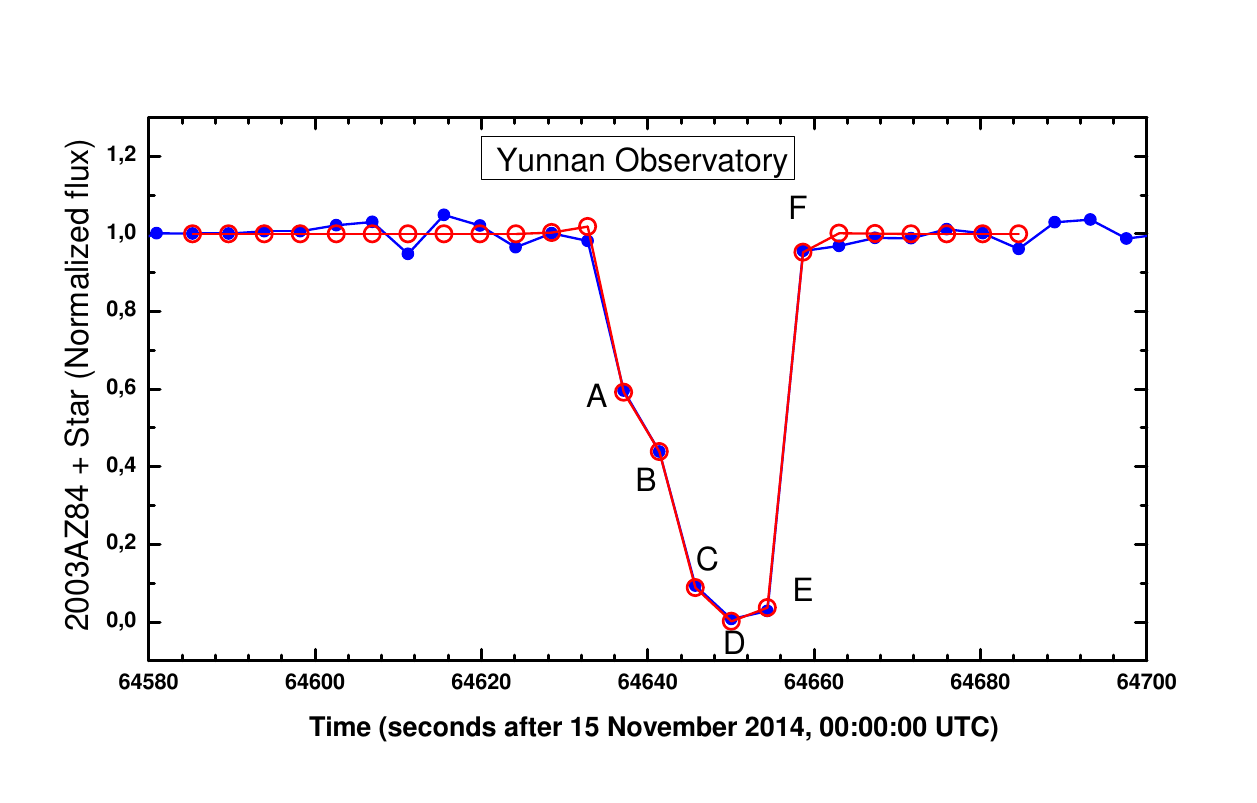}
\caption{%
Expanded view of the Yunnan light curve shown in Fig.~\ref{fig:lightcurves}, 
showing the gradual drop at ingress.
The model (red lines and circles) has been generated using the two possible
topographic solutions displayed in Fig.~\ref{fig:chasm_detail} 
(providing identical fit to the data).
}%
\label{fig:yunnan_fit} 
\end{figure}
\end{center}

Fig.~\ref{fig:yunnan_fit} shows the gradual disappearance of the star during ingress
at the Yunnan station.
It took more than 8~seconds for the stellar flux to go from its unocculted level to complete disappearance.
This corresponds to a displacement of the stellar image of more than 80~km along the local limb.
Since this is the only instance of gradual dimming observed during this event, 
it must come from a localized feature on the limb and cannot stem from a star duplicity.
Various causes might explain this behavior,
the most plausible one being a topographic feature revealed by the grazing geometry of the chord
(Fig.~\ref{fig:elipse_fit_2014}).

Beyond a topographic origin, only exotic explanations may be envisaged, 
like the presence of a local dust cloud lingering over the surface, or
a local atmosphere that refracts the stellar rays.
The drops of signal observed during intervals A and B in Fig.~\ref{fig:yunnan_fit} would imply
average tangential optical depths of $\tau_A = 0.53$ and $\tau_B = 0.83$, respectively, for dusty material
(assuming that point C is then be the usual star disappearance behind the solid limb). 
This would imply in a very dense local dusty cloud with length $\sim 80$~km and height $\sim10$~km.
%
An intense cometary-like activity would be required to create such cloud, 
an unlikely process so far away  from the Sun ($\sim$~45 AU), 
and such an activity has not been reported before for 2003~AZ$_{84}$.
 
On the other hand, \cite{Ortiz12} analyzed  the effects of a localized atmosphere during an occultation.
Typical surface pressures of a few microbars are required to create the feature of Fig.~\ref{fig:yunnan_fit}.
A stationary local atmosphere could result from a delicate balance between local sublimation and 
recondensation farther away on the surface.
To analyze further this scenario, it is necessary to know the pole position of the object in order to
determine the sub-solar latitude along the limb (where sublimation would be favored). 
Moreover, an albedo map of 2003~AZ$_{84}$ would be useful to identify warmer/cooler terrains
where sublimation/condensation could be favored.
These pieces of information are missing right now, preventing a more detailed discussion of the local atmosphere hypothesis.


Returning to the topographic feature explanation, we note that each integration interval lasts for 3~seconds
(Table~\ref{tab_circum_2014}), corresponding to about 28~km traveled by the star parallel to the limb,
while the complete duty cycle (4.3~s) corresponds to about 40~km.
In that context, there is an infinity of solutions for the limb structure that fit equally well the observed light curve.
For instance, the star may disappear and re-appear several times behind local reliefs during a single acquisition interval.

\begin{center}
\begin{figure}[!htb]
\centering
\includegraphics[scale=1.2]{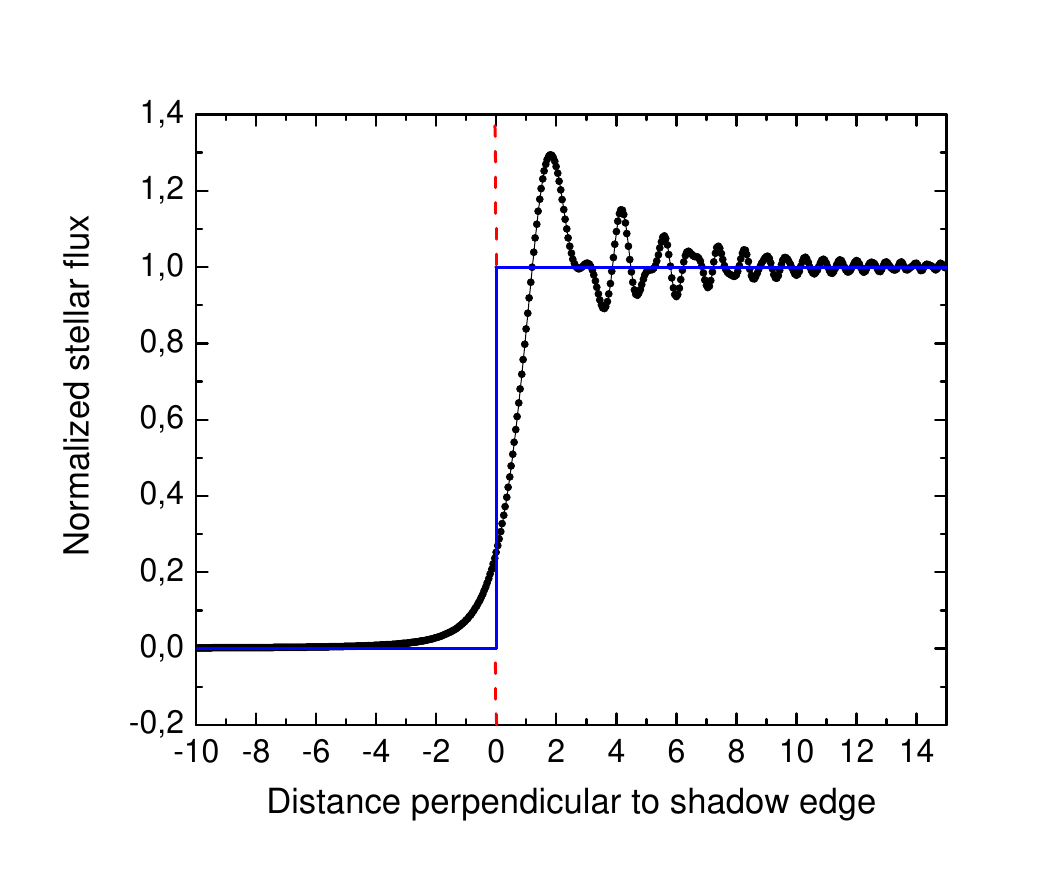}
\caption{%
Black line and dots: 
the Fresnel diffraction pattern perpendicular to a shadow edge observed at $6.67 \times 10^{9}$~km (44.5 AU), 
corresponding to the geocentric distance of 2003~AZ$_{84}$ on November 15, 2014.
The blue line is the abrupt edge profile (located along the red dashed line) that would be obtained in the limit of geometric optics 
and with a point-like star.
Moreover, the pattern has been averaged over a spectral interval of 0.3~$\mu$m in the visible domain
to account for the broadband response of the CCD at the Yunnan station.
}%
\label{fig:chasm_limb_profile}
\end{figure}
\end{center}

We examine here two extreme, simple solutions that provide possible ranges for the width and depth of the limb feature.
%
One solution (Solution~1) assumes that the observed light curve is caused by 
a chasm with vertical, abrupt walls (that is, perpendicular to the global fitted limb displayed in Fig.~\ref{fig:elipse_fit_2014}), 
while Solution~2 assumes that the topographic features is smooth with very shallow slopes, 
corresponding to a local limb that is always parallel to the global limb.
In both solutions, we have to convolve the sharp shadow edge profile by 
the Fresnel diffraction pattern, 
the stellar diameter and 
the integration time.
The Fresnel pattern corresponding to the November 15, 2014 event is displayed in Fig.~\ref{fig:chasm_limb_profile}. 
As  the projected stellar diameter is estimated to be 0.2~km, it is negligible compared to the Fresnel scale, $\sim 1.5$~km
(see Section~\ref{sub:OCCULTATION-TIMING}).
Thus the occultation light curve is dominated by Fresnel diffraction, and of course, by the large integration
time (3~s) at the Yunnan station. 

In the Solution~1 scenario, the star first disappears behind the global limb,
then it re-appears in the chasm during interval B, and disappear again during interval C.
During intervals D and E, the star is completely hidden behind the body, 
and it re-appears from behind the global limb near the start of interval F.
The locations of the walls are then adjusted so as to reproduce the observed fluxes (Fig.~\ref{fig:yunnan_fit}).
%

\begin{center}
\begin{figure}[!htb]
\centering
\includegraphics[scale=1.2]{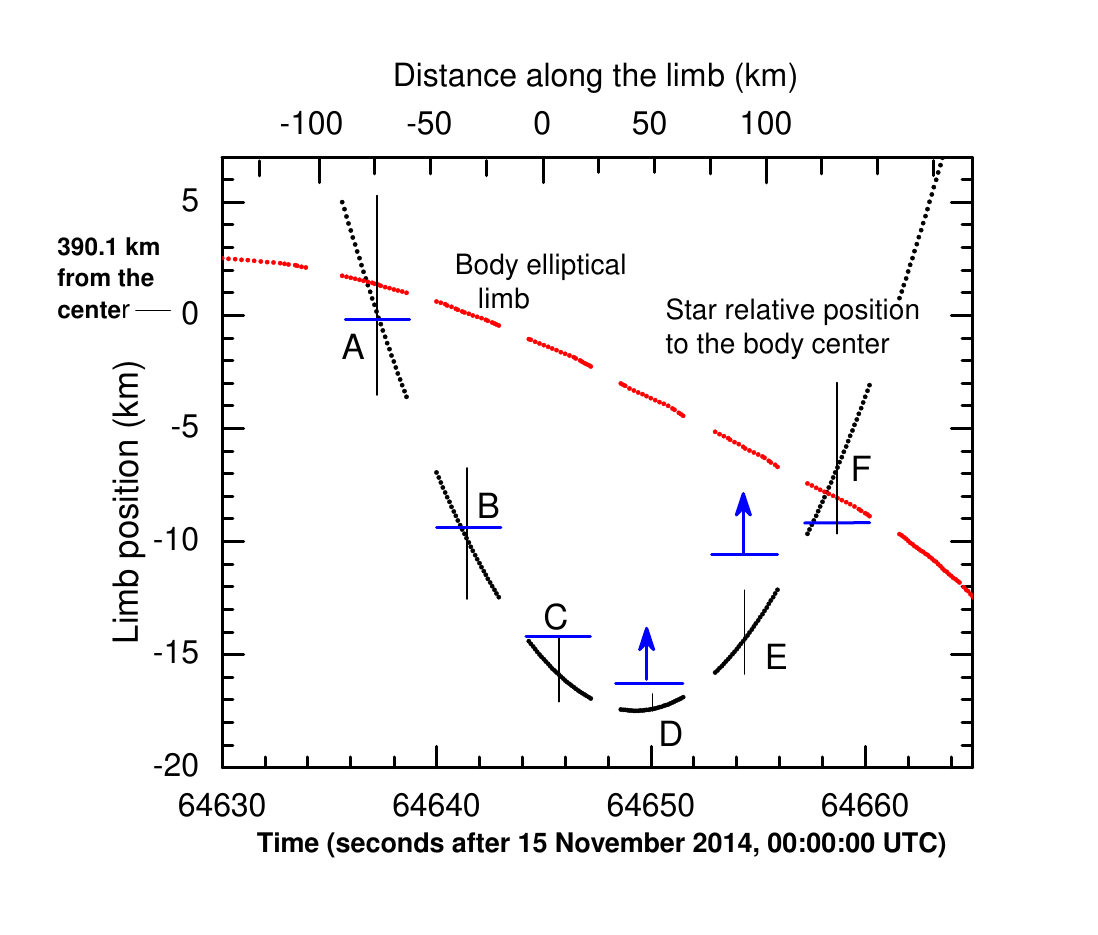}
\caption{%
Modeling the shallow depression solution (Solution~2 in the text).
The black dots along the parabola show the star position relative to 2003~AZ$_{84}$'s center
at 0.1~s time steps, inside each integration interval (solid blue lines).
The gaps correspond to the read out time of each image, with no data acquired. 
The origin of the vertical scale has been chosen arbitrarily so as to correspond to a radial
distance of 390.1~km (close to the distance corresponding to the mid-time of interval A)
The solid vertical lines are the mid-time of each exposure and the solid red lines are the limb radial
distances provided by the global elliptical fit. 
The vertical position of each blue line marks the limb offset for each 
acquisition interval that reproduces the observed normalized flux of Fig.~\ref{fig:yunnan_fit}. 
The blue arrows on intervals D and E indicate that only a lower limit for the local limb offset
could be derived (as no signal from the star was recorded during those intervals). 
The upper horizontal axis shows the distance traveled by the star relative to the limb,
with arbitrary origin corresponding to interval C.
See text for more details.
} 
\label{fig:chasm_limb_fit}
\end{figure}
\end{center}

\begin{center}
\begin{figure}[!htb]
\centering
\includegraphics[scale=1.2]{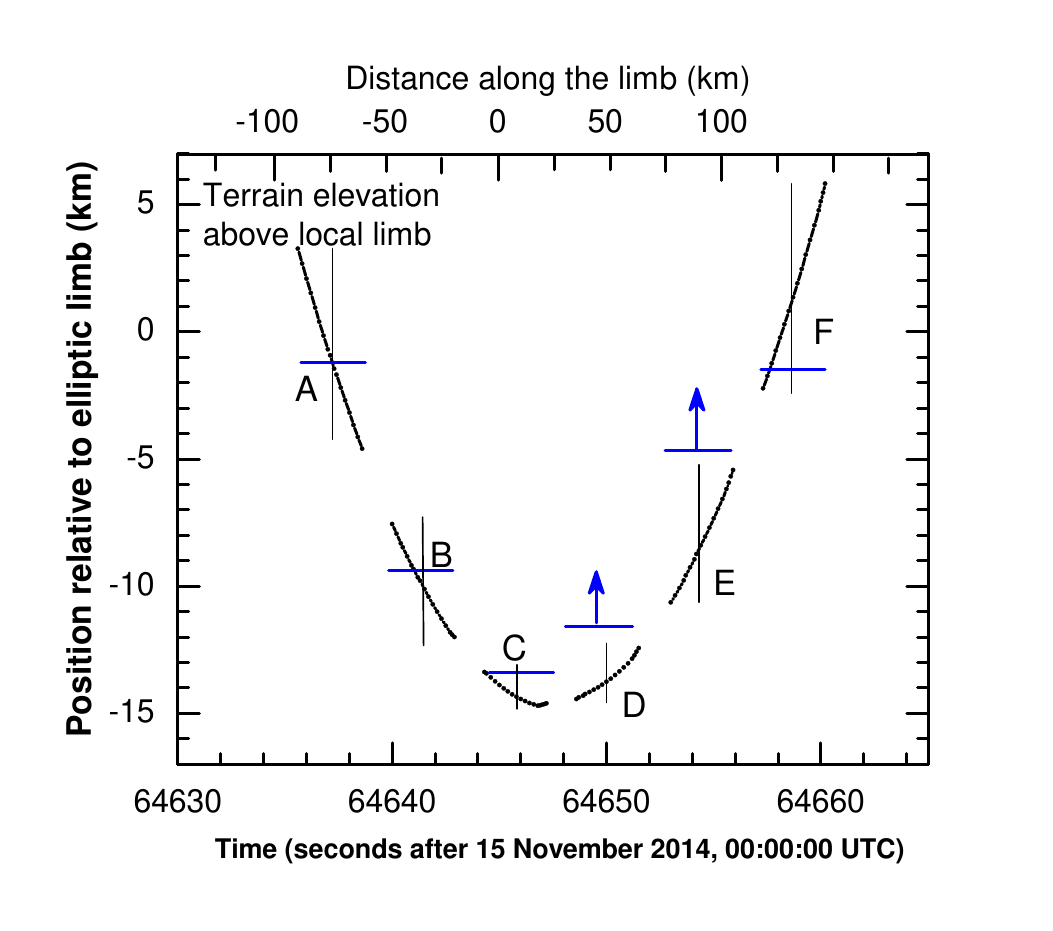}

\caption{%
Same as Figure \ref{fig:chasm_limb_fit}, but showing the difference in height between 
the blue lines (and black dots) and the red lines from Figure ~\ref{fig:chasm_limb_fit}.
This provides the average elevation of the terrain in each interval with respect to the global elliptical limb,
see also Fig.~\ref{fig:chasm_detail}. 
}%
\label{fig:chasm_limb_fit_2} 
\end{figure}
\end{center}

Solution~2 is a bit more complex to implement, as the star velocity relative to the body center
varies significantly during each acquisition interval.
To generate a synthetic light curve, we calculate the position of the star relative to the body center
at regular time steps of 0.1~second inside each integration interval (black dots in Fig.~\ref{fig:chasm_limb_fit}).
Using the Fresnel pattern of Fig.~\ref{fig:chasm_limb_profile}, 
we can generate the synthetic flux by averaging the theoretical fluxes calculated at each 0.1~s step inside a given interval.
The free parameter in this calculation is the radial offset of the local limb
(corresponding to the blue lines in Fig.~\ref{fig:chasm_limb_profile}), 
adjusted so as to fit the observed profile.
This offset provides in turn the difference between the positions of the local limb and that of the global limb 
(red lines in Fig.~\ref{fig:chasm_limb_fit}), 
and eventually the average elevation of the terrain in each interval (Fig.~\ref{fig:chasm_limb_fit_2}). 

Fig.~\ref{fig:chasm_detail} summarizes our results. 
The first panel sketches our inferred chasm structure (Solution~1)
with a width of $22.6 \pm 0.4$~km, 
while only a lower limit of about 8~km  
can be given for its depth.
The second panel (Solution~2) reveals a general depression that reaches more than 13~km in depth and 
extends over more than 80~km along the limb.

These structures can be compared with topographic features recently revealed on Pluto's and Charon's surfaces
during the New Horizons flyby \citep{stern15,nimmo16}. 
On both bodies, features reaching depths of up to 10~km are observed, some of them extending
over tens of kilometers on the surface. This supports our finding of a chasm or depression on 2003~AZ$_{84}$'s
surface, as this body is smaller (semi-major axis  radius $a < 500$~km, see previous section) 
than Pluto and Charon  (radii 1188~km and 606~km respectively, see \citealt{nimmo16}), 
and thus can in principle sustain comparable or deeper topographic features.


\section{CONCLUSIONS}
\label{sec:CONCLUSIONS}

We have combined data from four stellar occultations by the plutino object 2003~AZ$_{84}$ observed in 
2011 (January 8), 
2012 (February 3), 
2013 (December 2) and 
2014 (November 15), 
two of them being single-chord (2011 and 2013) and the other two being multi-chord events (2012 and 2014).

\begin{figure*}[!htb]
  \centering
\includegraphics[height=7.5cm]{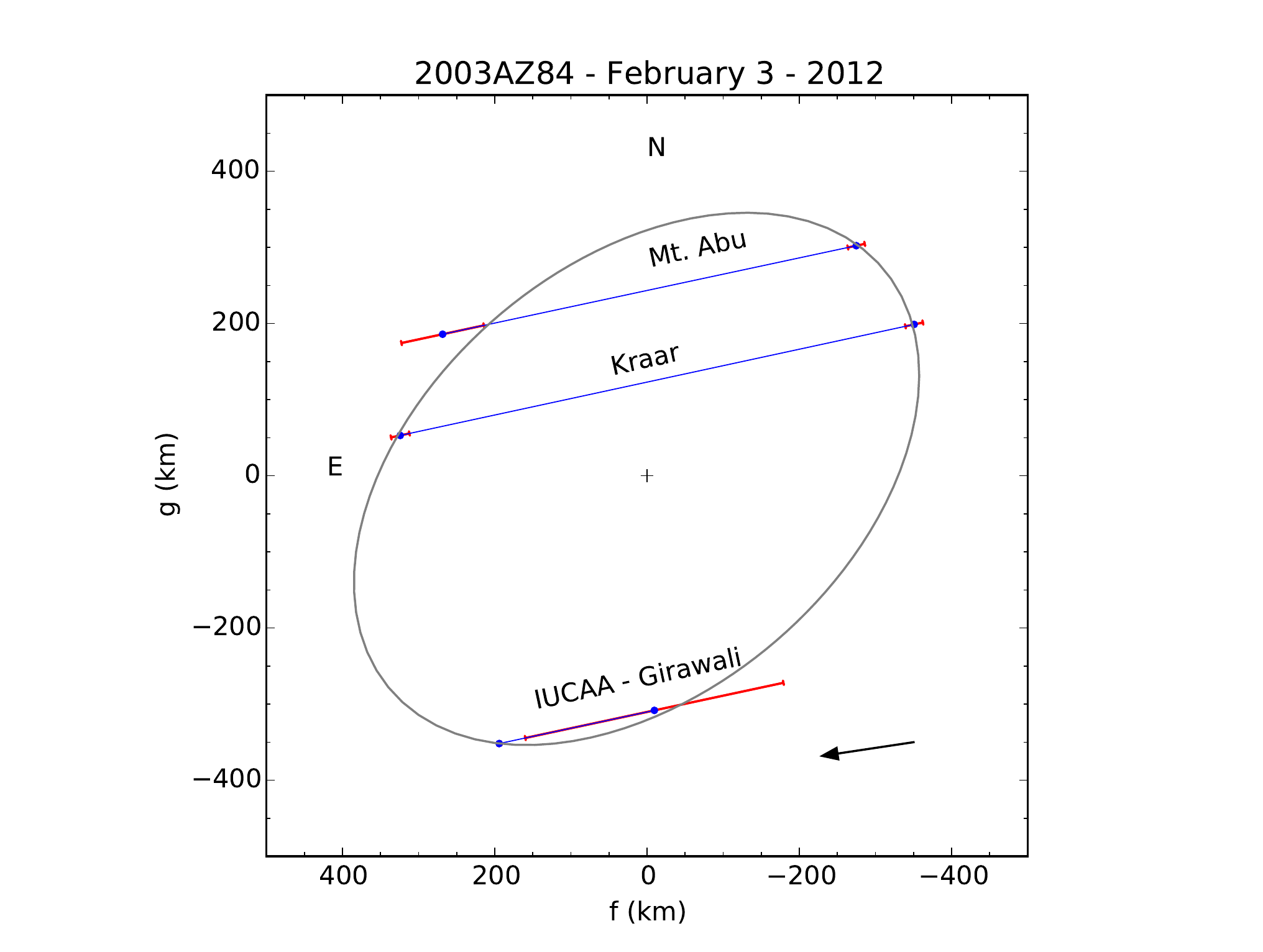}
\caption{%
\normalsize
Blue lines: 
the occultation chords showing the stellar motion (from right to left, see arrow) relative to 2003~AZ$_{84}$'s 
fitted center, as observed from the three station of February 3, 2012.
The red segments indicate the 1-$\sigma$ error bars on each chord extremity, 
derived from the time uncertainties provided in Table~\ref{tab_ingress_egress}.
Here, the star offsets $(f,g)$ in right ascension and declination, respectively, 
have their origin arbitrarily fixed at the body fitted center (cross) and are expressed in kilometers.
Black line: 
the best elliptical limb fit, whose geometric parameters are listed in Table~\ref{tab_results}.
Label N (resp. E) indicates local celestial north (resp. east).
Besides the chords shown here, no secondary events were detected, 
as could be caused by satellites or rings.
}%
\label{fig:elipse_fit_2012}
\end{figure*}

\begin{figure*}[!htb]
\centering
\includegraphics[height=7.5cm]{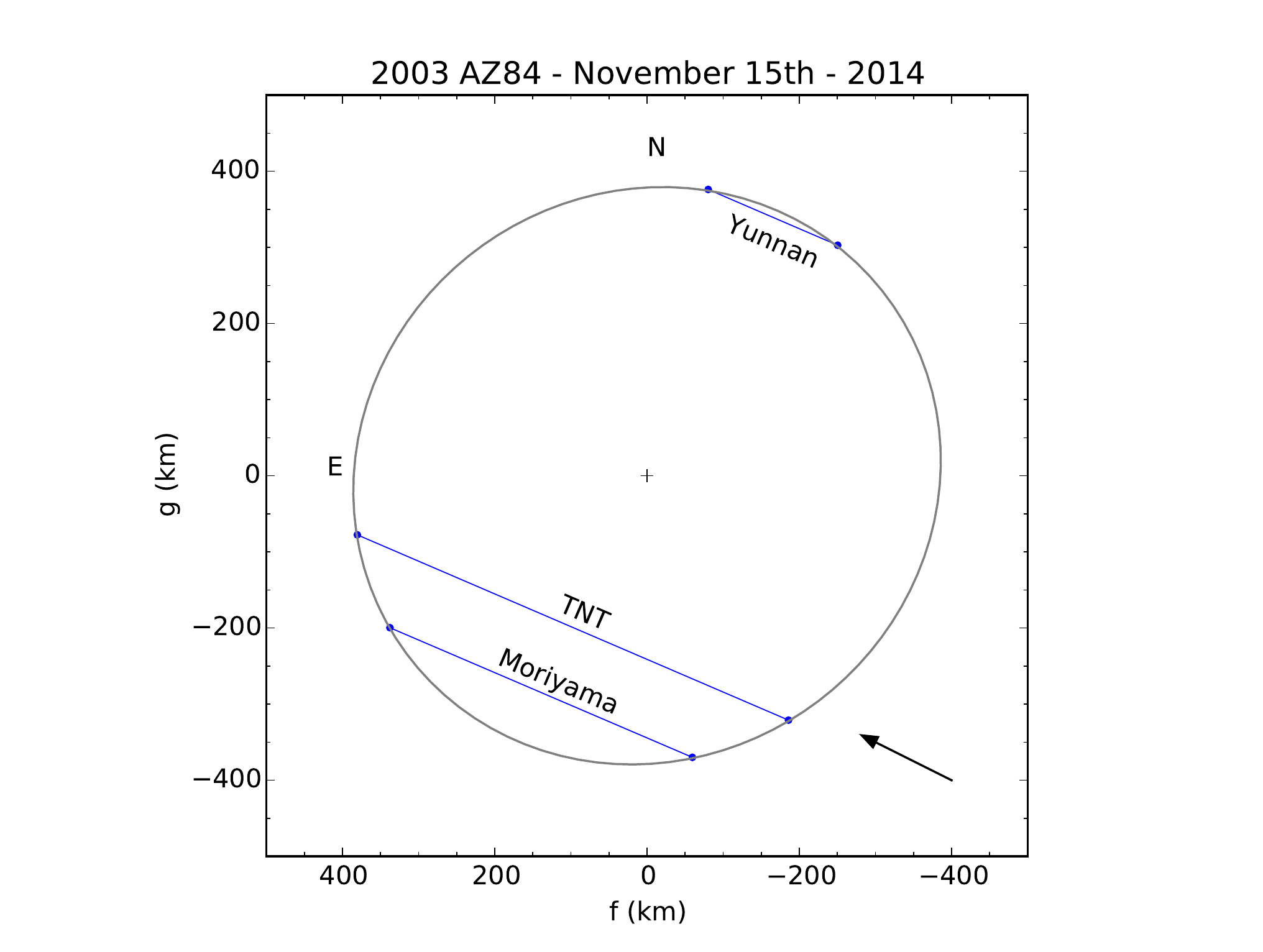}
\caption{%
\normalsize
Same as Figure \ref{fig:elipse_fit_2012}, but for the November 15, 2014 event.
}%
\label{fig:elipse_fit_2014}
\end{figure*}

The multi-chord events provide different limb shapes (Figs.~\ref{fig:elipse_fit_2012} and \ref{fig:elipse_fit_2014})
that we interpret as being due to a Jacobi ellipsoid that has been observed in 2012 and 2014 at two different rotation angles.
This Jacobi solution has 
a semi-major axis $a= 470 \pm 20$~km, 
axis ratios $b/a= 0.82 \pm 0.05$ and $c/a= 0.52 \pm 0.02$
(corresponding to $b=383 \pm 10$~km and $c=245 \pm 8$~km) and
density $\rho=0.87 \pm 0.01$~g~cm$^{-3}$.
%
%
The equivalent radius $R_{\rm eq} = 386\pm6$~km that we derive for  2003~AZ$_{84}$ (in 2012 and 2014) provides 
a visible geometric albedo of $p_V= 0.097 \pm 0.009$.
Both values are consistent with those derived from thermal measurements,
$R_{\rm eq} = 364_{-33}^{+31}$~km and $p_V = 0.107_{-0.016}^{+0.023}$ \citep{mommert12}, 
but  with higher accuracy.
%

\begin{figure}[!htb]
  \centering
\includegraphics[height=10cm]{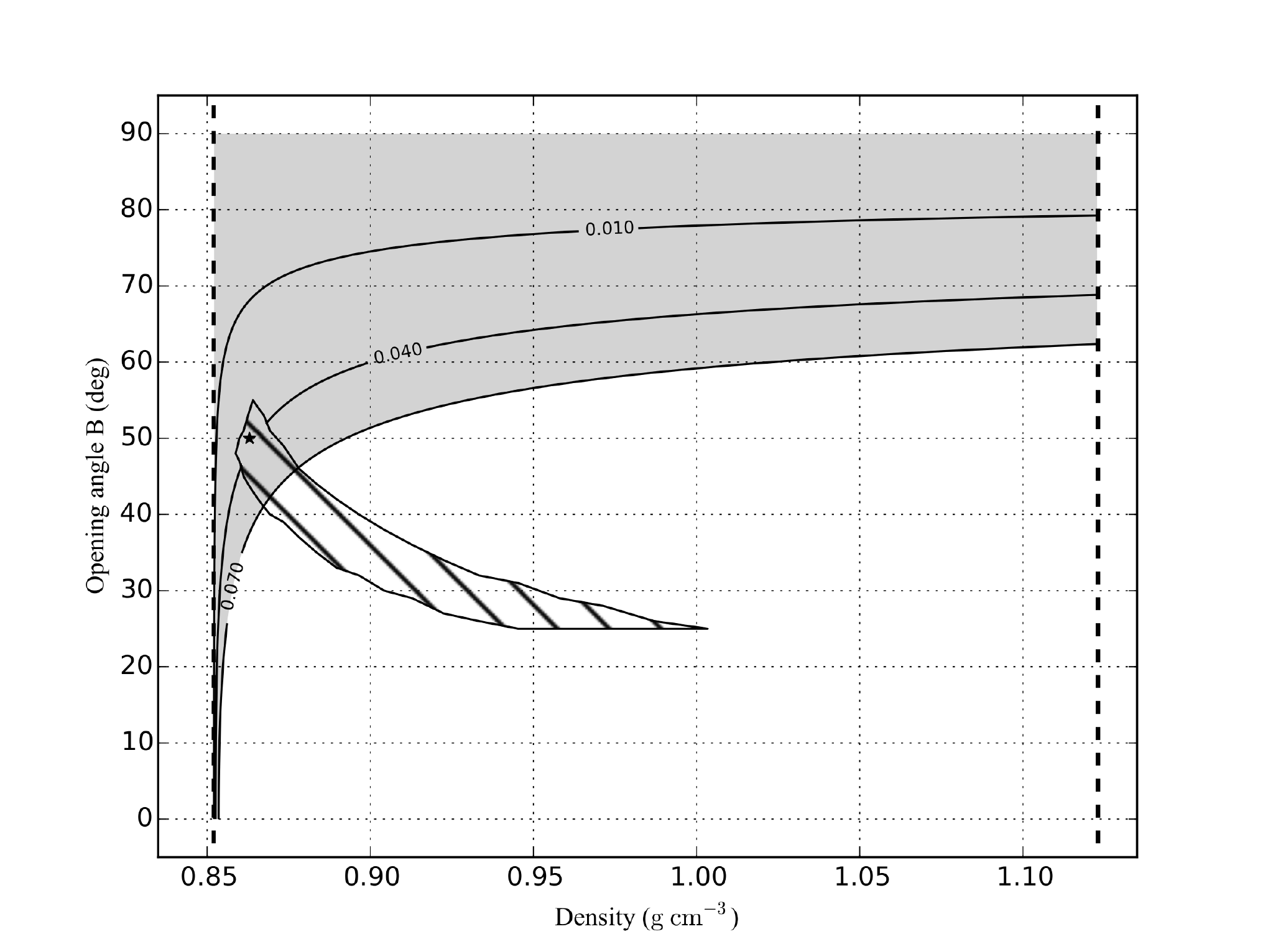}
\caption{%
The hatched band shows the range (1-$\sigma$ level) of possible opening
angles $B$
($B=0^\circ$ and $B=90^\circ$ corresponding to equator-on and pole-on
orientations, respectively)
as a function of density for a Jacobi ellipsoid with rotation period $P =
6.75$~h,
using the limb solutions shown in Figs.~\ref{fig:elipse_fit_2012} and
\ref{fig:elipse_fit_2014}.
\textbf{The dashed vertical lines at $ \rho=0.853$ and $\rho=1.12$~g~cm$^{-3}$} are
the theoretical limits for the Jacobi solutions.
The solid black lines are the loci of constant rotational light curve
amplitude
($\Delta m = 0.01, 0.04, 0.07$ mag) for a Jacobi ellipsoid, using Eq.~5 of
\cite{bin89}.
The shaded region correspond to the constraint $\Delta m < 0.07$ provided
by observations, see text.
The intersection of the band with the shaded region provides a density
$ \rho \sim 0.87 \pm 0.01$~g~cm$^{-3}$ and opening angle $ B \sim
48^{\circ} \pm 7^{\circ}$.
The star symbol defines the solution illustrated in
Fig.~\ref{fig:representative_jacobi_solution}.
}%
\label{fig:jacobi_opening_angle_vs_density}
\end{figure}

\begin{figure}[!htb]
\centering
\includegraphics[height=7.1cm]{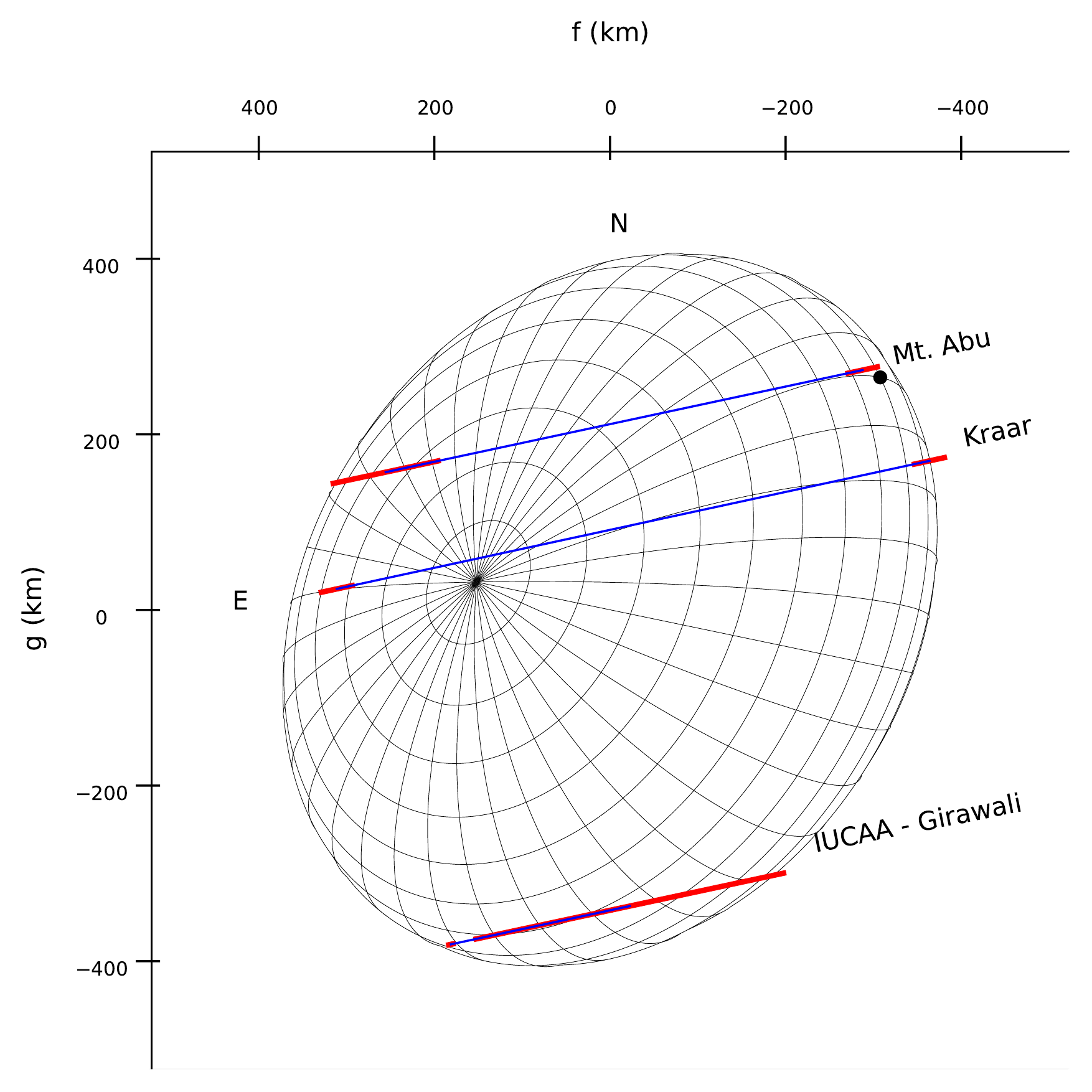}
\includegraphics[height=7cm]{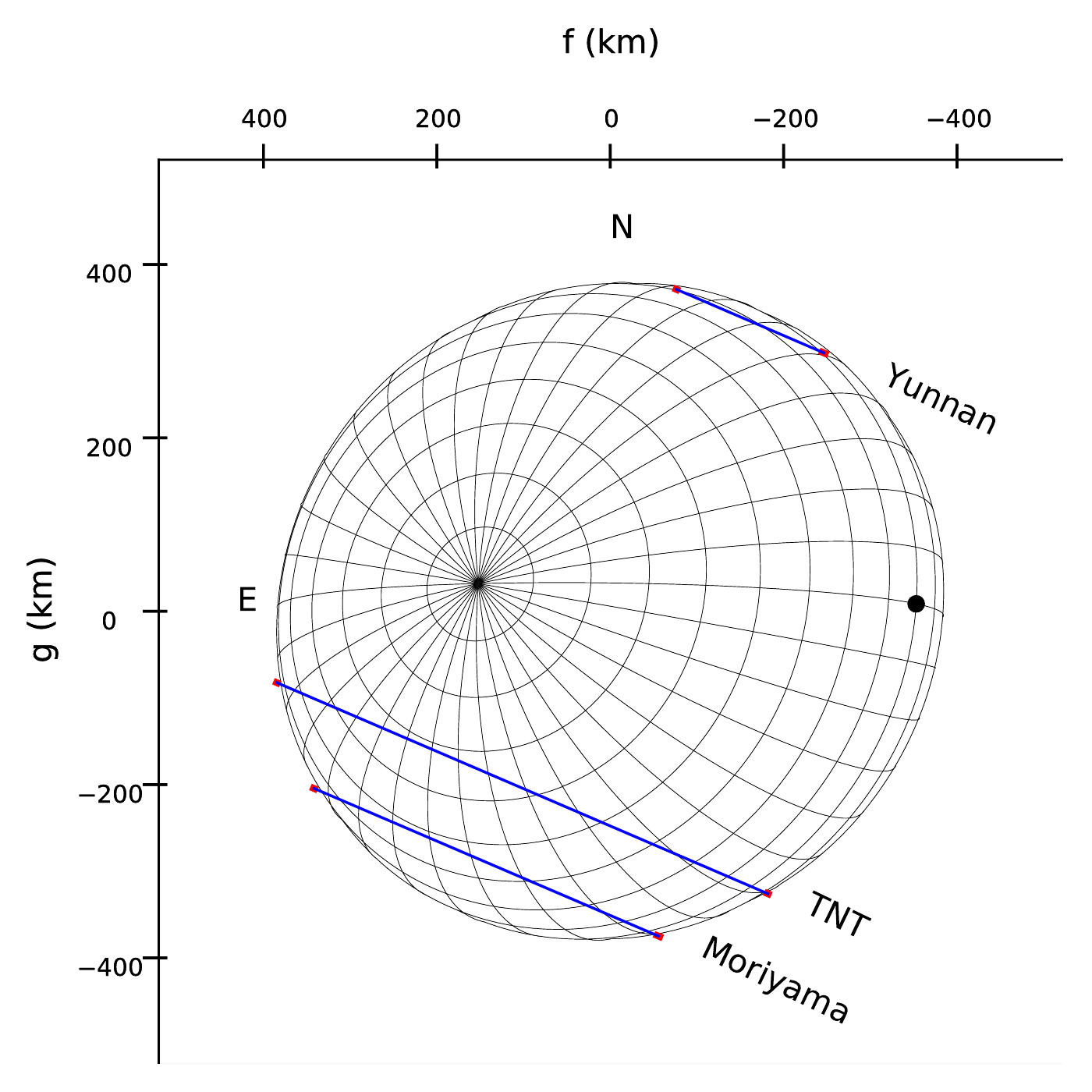}
\caption{%
A possible Jacobi ellipsoid solution that fits both 2012 and 2014 occultation results.
It has semi-major axes
$a \times b \times c = 456 \times 382 \times 242$~km,
opening angle $B=50^{\circ}$ and density $\rho=0.86$~g~cm$^{-3}$,
corresponding to the dot in Fig.~\ref{fig:jacobi_opening_angle_vs_density}. 
The \textbf{position angle of the pole is $P_p = 78^{\circ}$} (not to be confounded with the \textbf{position angle of the limb $P_L$, see text}).
The black dot marks the intersection of the longer axis $a$ with the surface of the body.
Left panel: fit to the 2012 event, using a rotation angle $ Q=45^{\circ}$. 
Right panel: fit to the 2014 event, using a rotation angle $ Q=10^{\circ}$. 
See text for details.
}%
\label{fig:representative_jacobi_solution}
\end{figure}

Besides yielding accurate sizes, these
occultations also constrain the shape of the body.
Some words of caution should be mentioned here, as we have only two multi-chord events at hand,
and some unaccounted timing errors might be present in some of our data.
Consequently, more occultations are now required to confirm our Jacobi solution.
This said, 
we have devised a general method, summarized in Fig.~\ref{fig:jacobi_opening_angle_vs_density},
that combines several multi-chord occultations and the amplitude of rotation derived from ordinary light curves to eventually pin down
(or rule out) a Jacobi solution for the body.
This method can be employed for analyzing future occultations by 2003~AZ$_{84}$
and  other TNOs.
Note that as more observations are added, this method will put stringent constraints on
the pole position, which in turn can be compared with the orientation of the satellite's orbit, when available.
The orbital period of the satellite will also provide 2003~AZ$_{84}$'s mass,
which in turn can be compared with the mass derived from the occultations, as described above.

The shape and density of 2003~AZ$_{84}$ can be compared with that of other elongated
TNOs like Haumea and Varuna, which have respective rotation periods of 3.9~h and 6.4~h \citep{thirouin16,Mueller15}.
Using observed rotation light-curves, \cite{lac07} infer values of 
$\rho \sim2.6$ and 
$\rho \sim 1.0$~g~cm$^{-3}$ for 
Haumea's and Varuna's densities, respectively, 
with corresponding axis ratios 
$(b/a \sim 0.83, c/a \sim 0.53)$ and 
$(b/a \sim 0.73, c/a \sim 0.49)$ for each object.
Thus, 2003~AZ$_{84}$ should resemble Haumea and Varuna in terms of shape,
with a density that is close to that of Varuna,
adding one more object to the family of very elongated TNOs.

\begin{center}
\begin{figure}[!htb]
\centering
\includegraphics[height=68mm,trim=0 0 0 0]{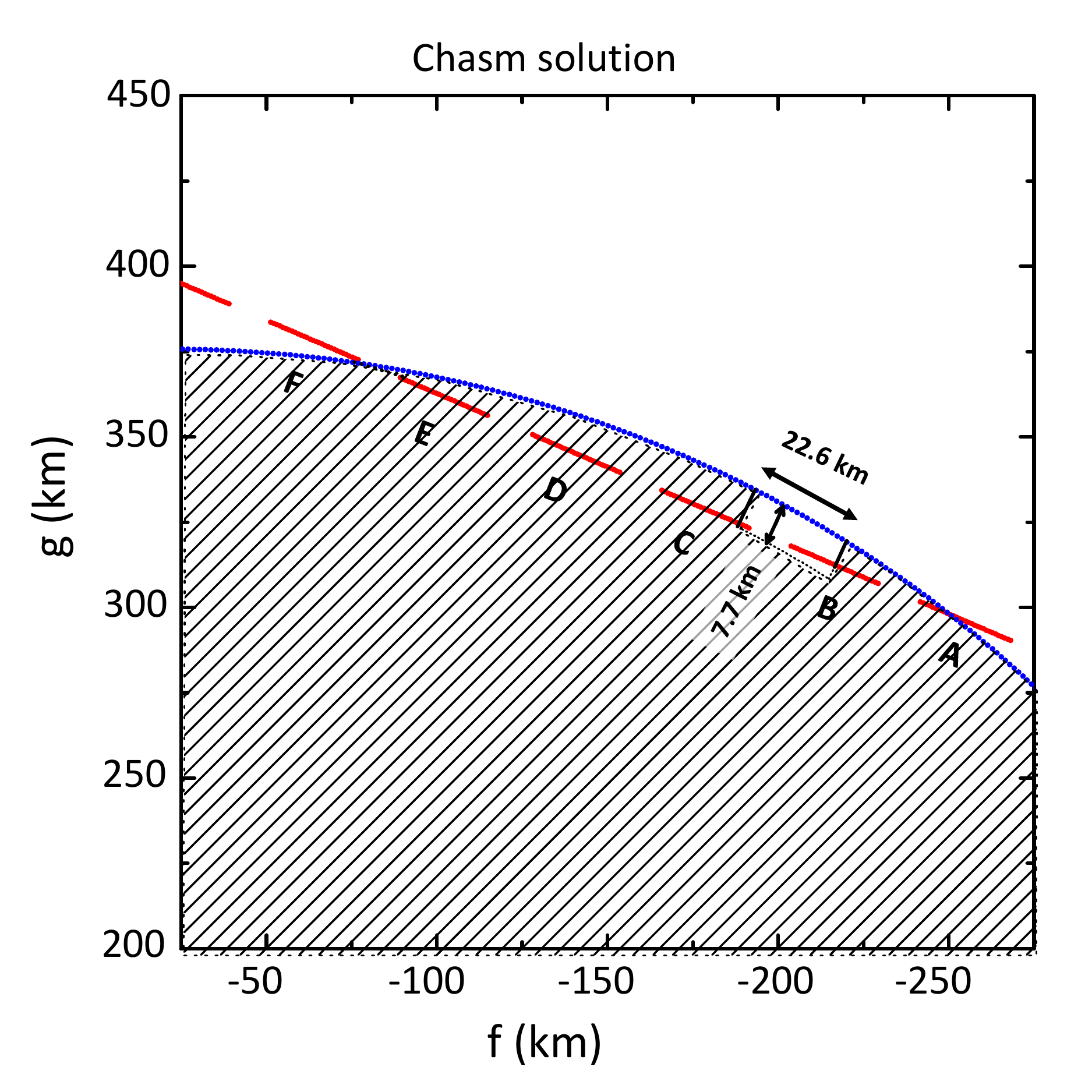}
\includegraphics[height=68mm,trim=0 0 0 0]{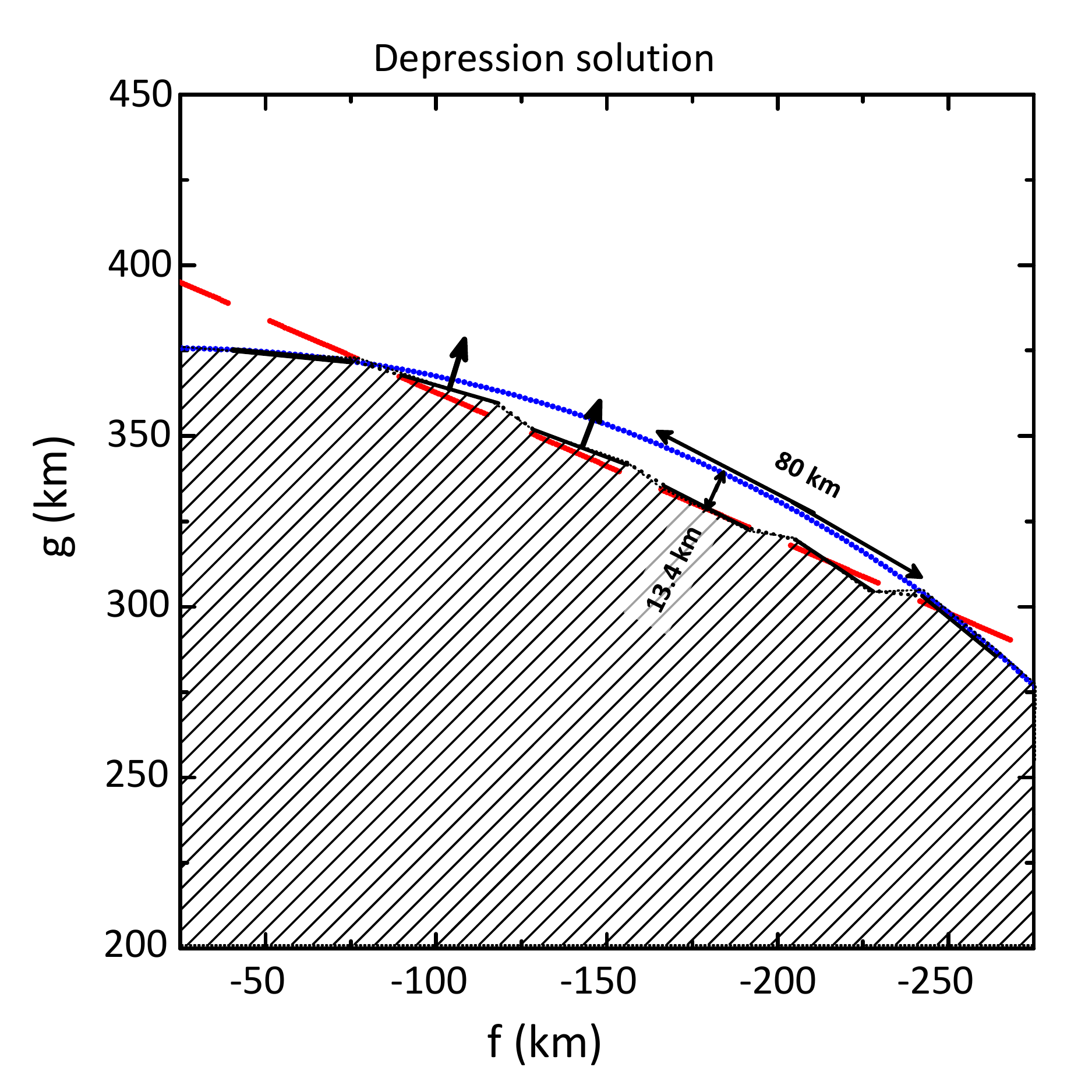}
\caption{%
Two possible limb solutions that can explain the gradual ingress at the Yunnan site 
during the November 15, 2014 occultation (Fig.~\ref{fig:yunnan_fit}).
In each panel, the blue line is the global fitted limb (see Fig.~\ref{fig:elipse_fit_2014})
and the red segments show the star motion during each exposure interval.
Left panel:
chasm solution (Solution~1 in the text) with vertical walls. 
The width of the chasm is $22.6 \pm 0.4$~km but only a lower limit of 7.7~km for its depth 
can be derived from the data.
Right panel:
depression solution (Solution~2).
The black segments are the positions of the average limb in each integration interval.  
The black arrows on segments D and E indicate that they are the deepest possible
levels for the limb that can be derived from the data
(since the star is not visible during those integration intervals, see Fig.~\ref{fig:yunnan_fit}).
As a consequence, a lower limit for the horizontal extension of the depression is 80~km, 
while its depth 13.4~km.
See text for details.
}%
\label{fig:chasm_detail} 
\end{figure}
\end{center}

Another noteworthy result obtained here is the detection of a topographic feature at the surface of
2003~AZ$_{84}$, a premi\`ere if we exclude Pluto and Charon.
Due to the lack of time resolution in our data, we cannot discriminate between 
an abrupt chasm of width $\sim 23$~km and depth $> 8$~km, 
a depression with shallow slopes, width of $\sim 80$~km and
depth of $\sim 13$~km, or other intermediate solutions (Fig.~\ref{fig:chasm_detail}).
Again, more occultations are required to pin down the structure of this kind of features. 
Note, however, that gradual star dis- or re-appearances occur if the occultation
chord is at less than typically 10~km from the average limb of the body.
In the Gaia era, star positions will be given at accuracies better than 1~mas.
In this context, successful occultations will themselves improve the ephemeris of the body, 
eventually providing predictions of new occultations with accuracies as good as $\sim 1$~mas.
This corresponds to about 30~km at 2003~AZ$_{84}$'s distance, or about 10~km for typical Centaur objects.
This will turn grazing occultations by remote bodies from a hopeless task to a routine method
that will reveal rims, chasm, craters (with irregular edges), basins, etc..., thus fostering geological studies of those objects.


{\bf Acknowledgments.}
We acknowledge support from the French grants 
``Beyond Neptune"  ANR-08-BLAN-0177 and
``Beyond Neptune II"  ANR-11-IS56-0002.
Part of the research leading to these results has received funding from
the European Research Council under the European Community's H2020
(2014-2020/ERC Grant Agreement no. 669416 ``LUCKY STAR").
A. Dias-Oliveira thanks the support of the following grants: CAPES (BEX 9110/12-7) FAPERJ/PAPDRJ (E-45/2013).
R. Leiva acknowledges support from CONICYT-PCHA/Doctorado Nacional/2014-21141198.
R. Duffard, J.~L. Ortiz and  P.  Santos-Sanz 
have received funding from the European Union's Horizon 2020 
Research and Innovation Programme, under Grant Agreement no 687378.
R. Vieira-Martins thanks the following grants: CNPq-306885/2013, Capes/Cofecub-2506/2015, Faperj: PAPDRJ-45/2013 and E-26/203.026/2015.
M. Assafin thanks the CNPq (Grants 473002/2013-2 and 308721/2011-0) and FAPERJ (Grant E-26/111.488/2013).
J. I. B. Camargo acknowledges a CNPq/PQ2 fellowship 308489/2013-6.
This work has made use of data obtained at the Thai National Observatory on Doi Inthanon, operated by NARIT.
A. Maury acknowledges the use of the C. Harlingten telescope of the Searchlight Observatory Network
Funding from Spanish grant AYA-2014-56637-C2-1-P is acknowledged, as
is the Proyecto de Excelencia de la Junta de Andalucía, J. A.
2012-FQM1776.
A. Thirouin acknowledges funding from Lowell Observatory. 
G.B.R. is thankful for the support of the CAPES (203.173/2016) and FAPERJ/PAPDRJ (E26/200.464/2015 - 227833) grants.

\vspace{0.5cm}

\clearpage

\begin{deluxetable}{ccc}
\tabletypesize{\scriptsize}
\tablecaption{%
Astrometric observations for the 2012 event prediction.
\label{tab_astrometry_2012}
}%
\tablewidth{0pt}
\tablehead{
\colhead{Site}            & \colhead{Date}         & \colhead{Telescope}  \\
\colhead{IAU code}        & \colhead{mm/dd/yyyy}   &                      \\
}
\startdata
Pico dos Dias Observatory & 09/22/2011            & 0.60 m                \\
 874                      &                       & Boller \& Chivens     \\
\hline
La Hita Observatory      & 01/18/2012 to         & 0.77 m                \\
I95                      & 01/20/2012            &                       \\
\hline 
San Pedro de Atacama Observatory\footnotemark[1] & 01/16/2012 to         & 0.4 m                \\
I16                     & 01/25/2012            &     IAA-ASH2                  \\
\hline
Observatorio Astronomico de Cala d'Hort & 01/16/2012 to             & 0.5 m                \\
C85                     & 01/25/2012\footnotemark[2]    &                   \\
\hline
Pic du Midi Observatory\footnotemark[1] & 01/17/2012             & 1.0 m                \\
586                     &                    &                   \\
\hline
\enddata
\begin{footnotesize}
\footnotemark[1] The star and TNO were observed in the same field of view, allowing for the correction in their relative position.
\footnotemark[2] Data obtained in two nights in this range.
\end{footnotesize}

\end{deluxetable}

\begin{deluxetable}{cccccc}
\tabletypesize{\scriptsize}
\tablecaption{%
Observation circumstances for the February 3, 2012 event.
\label{tab_circum_2012}
}%
\tablewidth{0pt}
\tablehead{
 & \colhead{Longitude (W)} & \colhead{Telescope} & \colhead{Expo. Time} & \colhead{Observer} & \colhead{Detection} \\
\colhead{Site} & \colhead{Latitude (N)} & \colhead{Camera} & \colhead{Cycle} & & \\
& \colhead{Altitude (m)} &  & (s) & &
}
\startdata
Mt. Abu & -72$^{\circ}$ 46' 47.18'' & 1.2 m              & 4.0  & T. Baug, Jinesh Jain & positive  \\
 (India)& 24$^{\circ}$ 39' 10.34'' & NICMOS             & 4.98 & T.Chandrasekhar   &      \\
        & 1680                     &                    &      & S. Ganesh      & \\
\hline
IUCAA\footnotemark[1]   & -73$^{\circ}$ 40' 00.0''  & 2.0 m              & 2.0  & V. Mohan & positive\\
Girawali (IGO)                 & 19$^{\circ}$ 05' 00.0''  & IFOSC             & 16.5 &   &    \\
(India)                              & 1005                     &                    &      &  &   \\
\hline 
Kraar& -34$^{\circ}$ 48' 45.76' &  0.41 m              & 3.    0  & I. Manulis & positive\\
(Israel)& 31$^{\circ}$ 54' 29.1''  &  SBIG                  & 4.4-5.0? & E. Ofek &        \\
        & 107                      & ST-8XME CCD                   &          & A. Gal-Yam       &  \\
\hline
Liverpool  Tel.   & 17$^{\circ}$ 52' 45.2'' & 2.00 m              & 1.5  & Robotic   & negative \\
Canary          & 28$^{\circ}$ 45' 48.8''  & RISE          & 1.5  & Telescope     &    \\
   (Spain)         & 2457                      &                    &      &     &  \\
\hline
Wise   & -34$^{\circ}$ 45' 48.0'' & 1.00 m              & 3.0  & S. Kaspi   & Clouded\\
(Israel)    & 30$^{\circ}$ 35' 45.0''  & Princeton Instrument CCD    & 4.5  & N. Brosch     &    \\
            & 875                      &                    &      &     &  \\
\hline
Nikaya  & -77$^{\circ}$ 43' 33.0''  & 0.36 m                                 & 4.0  & A. A. Sharma & negative \\
Observatory & 12$^{\circ}$ 36' 23.0''  & SBIG ST-8XME CCD    & 6.2 &   &    \\
(India) & 1005                          &                    &      &  &   \\
\hline
Maidenhead       & -00$^{\circ}$ 48' 58.2''  & 0.30 m          & 2.56  & T. Haymes & negative \\
(United Kingdom) & 51$^{\circ}$ 30' 23.8''   & WATEC 120N+     & 2.56  &           &          \\
                 & 75                        &                 &       &           &          \\
\hline
Paris Observatori      & 01$^{\circ}$ 50' 19.9''   & 0.21 m         & 2.56  & J. Lecacheux & negative \\
- Mobile Station -       & 46$^{\circ}$ 31' 08.2''   & WATEC 120N+     & 2.56  &              &          \\
 (France)             & 305                       &                 &       &              &          \\
\hline
Cabrills               & 02$^{\circ}$ 23' 07.3''   & 0.30 m           & 10    & R. Naves     & negative \\
(Spain)                & 41$^{\circ}$ 31' 11.3''   &  SBIG ST8-XME       & 13    &              &          \\
                       & 114                       &                 &       &              &          \\
\hline
Alicante               & -359$^{\circ}$ 33' 18.0''   &    0.30 m        & 3.0    & R. G. Lozano     & negative \\
(Spain)                & 38$^{\circ}$ 28' 33.0''   &    SBIG ST8-XME    & 5.0    &              &          \\
                       & 187.3                       &       bin 3x3          &       &              &          \\
\hline
Tourrette Levens       & 07$^{\circ}$ 15' 47.2''   & 0.335 m            & 5.0    & P. Tanga     & negative \\
(France)               & 43$^{\circ}$ 47' 22.2''   & Apogee Alta U1 CCD & 5.68    &              &          \\
                       & 385                       &                    &       &              &          \\
\hline
Bellinzona             & 09$^{\circ}$ 01' 26.5''   & 0.40 m            & 5.0    & S. Sposetti     & negative \\
(Switzerland)          & 46$^{\circ}$ 13' 53.2''   &SBIG ST8 CCD & 7.0    &              &          \\
                       & 260                       &                    &       &              &          \\
\hline
Berlin                 & 13$^{\circ}$ 28' 30.8''   & 0.50 m            & 1.28-2.56    & G. Wortmann     & negative \\
(Germany)              & 52$^{\circ}$ 29' 12.5''   &WATEC 120N        & 1.28-2.56   &              &          \\
                       &  41                       &                    &       &              &          \\
\hline
Berlin                 & 13$^{\circ}$ 39' 41.5''   & 0.28 m            & 1.28    & S. Andersson     & negative \\
(Germany)              & 52$^{\circ}$ 24' 35.4''   & MINTRON 12V6      & 1.28   &              &          \\
                       &  45                       &                   &       &              &          \\
\hline
Nonndorf               & 15$^{\circ}$ 14' 08.7''   & 0.254 m            & 2.56    & G. Dangl     & negative \\
(Austria)              & 48$^{\circ}$ 47' 13.5''   & Watec 120N      & 2.56   &              &          \\
                       & 593                       &                   &       &              &          \\
\hline
\enddata
\begin{footnotesize}
\footnotemark[1] Inter-University Centre for Astronomy and Astrophysics
\end{footnotesize}

\end{deluxetable}

\begin{deluxetable}{cccccc}
\tabletypesize{\scriptsize}
\tablecaption{%
Observation circumstances for November 15, 2014 event.
\label{tab_circum_2014}
}%
\tablewidth{0pt}
\tablehead{
 & \colhead{Longitude (W)} & \colhead{Telescope} & \colhead{Expo. Time} & \colhead{Observer} & \colhead{Detection} \\
\colhead{Site} & \colhead{Latitude (N)} & \colhead{Camera} & \colhead{Cycle} &  & \\
& \colhead{Altitude (m)} &  & (s) & &
}
\startdata
Thai National       & -98$^{\circ}$ 25' 56.06'' & 2.4 m              & 0.786187  & A. Richichi  & positive \\
Telescope  & 18$^{\circ}$ 34' 25.41''  & ULTRASPEC          & 0.801169  & P. Irawati     &    \\
(Thailand)           & 2457                      &                    &      &     &  \\
\hline
Yunnan   & -100$^{\circ}$ 01' 51.0''  & 2.4 m            & 3.0  & Z. Liying & positive\\
(China)  & 26$^{\circ}$ 42' 32.0''   & 1340$\times$1300          & 4.3164 & Q. Shengbang &     \\
         & 3193                      &  CCD Camera       &      & Z. Ergang   &  \\
\hline 
Moriyama& -135$^{\circ}$ 59' 23.8'' &  0.26  m            & 10.0      & Y. Ikari & positive \\
(Japan) & 35$^{\circ}$ 02' 59.1''  &  Apogee ALTA U6    & 11.65      &     &     \\
  CCD   & 105                      &                    &          &       &   \\
~\\
Moriyama&  &  0.254              & 08.53      & Y. Ikari & positive \\
(Japan) &   &  WAT-120(slow shutter 256Fr)            & 08.53      &     &     \\
 Video  &                     &                    &          &          \\
\hline
Inabe   & -136$^{\circ}$ 31' 24.7''  & 0.35                       &  08.53  & A. Asai & Observation \\
(Japan)  & 35$^{\circ}$ 10' 14.7''    & TGv-M (slow shutter 256Fr) &  08.53  & Hayato Watanabe &  finished before   \\
         & 187                     &                            &      &    &  actual event \\
\hline
Hamamatsu   & -137$^{\circ}$ 44' 23.0'' &   0.254                    &  08.53  & M. Owada & negative \\
(Japan)    &  34$^{\circ}$ 43' 070'' & WAT-120(slow shutter 256Fr) &  08.53               &          &     \\
 Video     &  17                      &                             &                 &          &        \\
\hline
Tarui      & -136$^{\circ}$ 29' 43.7'' &   0.35                    & 2.13  & Hiroyuki Watanabe & Observation \\
(Japan)    &  35$^{\circ}$ 23' 41.4'' & TGv-M(Slow shutter 64Fr) &  2.13               &          &   finished before  \\
 Video     &  105                      &                             &                 &          &   actual event     \\
\hline
Hitachi     & -140$^{\circ}$ 41' 08.9'' &   0.30                    &  4.27  & H. Tomioka & Obs. Taken \\
(Japan)    &  36$^{\circ}$ 38' 33.2'' & WAT-120N+RC (slow shutter 128Fr) &     4.27       &          & on wrong    \\
 Video     &  33                      &                             &                 &          &    area  \\
\hline
Maibara     & -136$^{\circ}$ 17' 40.3'' &   0.13                    &  4.27  & H. Yamamura & Observation \\
(Japan)    &  35$^{\circ}$ 19' 41.8'' & WAT-910HX(slow shutter 128Fr) &    4.27          &          &   finished before  \\
 Video     &  92                      &                             &                 &          &   actual event     \\
\hline
\enddata


\end{deluxetable}

\clearpage

\begin{deluxetable}{ccccc}
\tabletypesize{\scriptsize}
\tablecaption{%
Ingress and Egress times.
\label{tab_ingress_egress}
}%
\tablewidth{0pt}
\tablehead{\multicolumn{5}{c}{\textbf{January 8, 2011}}\\
~\\
\colhead{Site} & \colhead{Ingress (UT)} & \colhead{Error (s)} & \colhead{Egress (UT)} & \colhead{Error (s)} \\}
\startdata
~\\
San Pedro Atacama (SPACE)     & 06:29:48.30            & 0.80                & 06:30:10.00           & 0.8   \\
~\\
\hline\hline
~\\
\multicolumn{5}{c}{\textbf{February 3, 2012}}\\
~\\
Mt. Abu         & 19:45:13.89            & 0.70                & 19:45:35.95           & 2.5   \\
~\\
\hline
~\\
Kraar        & 19:47:39.20            & 0.70                & 19:48:06.15           & 0.68   \\
~\\
\hline
~\\
IUCAA Girawali  & 19:45:17.16\footnotemark[1] & 7.3                & 19:45:25.43           & 0.10   \\
~\\
\hline\hline
~\\
\multicolumn{5}{c}{\textbf{December 2, 2013}}\\
~\\
Rockhampton         & 14:52:48.44            & 1.40                & 14:53:28.17           & 1.0   \\
~\\
\hline\hline
~\\
\multicolumn{5}{c}{\textbf{November 15, 2014}}\\
~\\
Thai National Telescope          & 17:58:09.57            & 0.02                & 17:59:13.80           & $^{+0.04}_{-0.03}$   \\
~\\
\hline
~\\
Yunnan         & 17:57:21.65            & 0.13                & 17:57:36.95           & $^{+0.13}_{-0.14}$   \\
~\\
\hline
~\\
Moriyama (CCD) & 17:52:15.32            & 0.18                & 17:52:59.90           & 0.20   \\
~\\
\hline
~\\
Moriyama (Video)& 17:52:15.25           & $^{+3.24}_{-3.99}$  & 17:52:55.69           & $^{+4.40}_{-4.01}$   \\
~\\
\hline
\enddata

\begin{footnotesize}
\footnotemark[1] Whereas the ingress must happen between 19:45:09.88 and 19:45:24.43 UTC, it can be represented by the mid-time between
them with the error being the difference between the two extremes and the midpoint.

\end{footnotesize}


\end{deluxetable}

\begin{deluxetable}{ccc}
\tabletypesize{\normalsize}
\tablecaption{%
Physical parameters for 2003~AZ$_{84}$ from the 2012 and 2014 multi-chord events.
\label{tab_results}
}%
\tablewidth{0pt}
\tablehead{  \colhead{Solution} & \colhead{2012}           & \colhead{2014} \\}
\startdata
~\\
 Apparent semi-major axis $a'$ (km)   &  $426^{+26}_{-16}$          & $393 \pm 1$ \\ \\
 Equivalent radius $R_e$  (km)           & $346 \pm 6$                 &  $382 \pm 3$ \\ \\
 Projected oblateness  $\epsilon'$        &  $0.340^{+0.097}_{-0.086}$     &   $0.054 \pm 0.003$ \\ \\
 $f_c$ (km)\footnotemark[1]  & $6800\bla ^{+23}_{-17}$       &    $2862^{+2.1}_{-1.8}$ \\ \\
 $g_c$ (km)\footnotemark[1] & $2850\bla^{+14}_{-11}$        &   $336^{+4.0}_{-2.7}$ \\ \\
Position angle of the limb $P_L$ (deg)   &  $41^{+8}_{-5}$                  &   $36 \pm 4$ \\ \\
$\chi^2_{pdf}$                     & 0.73                                &     0.98 \\ \\
\hline
~\\
\multicolumn{3}{c}{Jacobi solution} \\ \\
\hline
~\\
Semi-major axis $a$ (km) &  \multicolumn{2}{c}{ $470 \pm 20$} \\ \\ 
~\\
Axis ratios &   \multicolumn{2}{c}{$b/a= 0.82 \pm 0.05$, $c/a = 0.52 \pm 0.02$} \\ \\ 
~\\
Density $\rho$ (g cm$^{-3}$)                    &  \multicolumn{2}{c}{ $0.87 \pm 0.01$} \\ \\
Visible geometric albedo $p_V$    &  \multicolumn{2}{c}{$0.097 \pm 0.009$}   \\ \\
\hline
\enddata
\begin{footnotesize}
\footnotemark[1]
The quantities $(f_c,g_c)$ are the offsets, expressed in km, in 
right ascension ($\Delta \alpha \cos(\delta)$) and 
declination ($\Delta \delta$) to be applied to 
2003~AZ$_{84}$'s ephemeris (NIMA~V1),
as deduced from the elliptical fits to the chord extremities shown in Figs.~\ref{fig:elipse_fit_2012} and \ref{fig:elipse_fit_2014},
using the star positions of Eqs.~\ref{eq:a_d_star_2012} and \ref{eq:a_d_star_2014}.
The NIMA ephemeris provides the following reference J2000 geocentric positions for the body:
$(07^{\rm h} 45^{\rm m} 54.79070^{\rm s}, +11^{\circ}12' 43.0422'')$, and
heliocentric distance $\Delta= 6.6209 \times 10^9$~km on February 3, 2012 at 19:45 UTC, and
$(08^{\rm h} 03^{\rm m} 51.28512^{\rm s}, +09^{\circ}57' 18.7956'')$,
$\Delta= 6.6634 \times 10^9$~km on November 15, 2014 at 17:55 UTC.
Applying the $(f_c,g_c)$ offsets tabulated here, we obtain the following corrected geocentric positions for 2003~AZ$_{84}$:
$ (07^{\rm h} 45^{\rm m} 54.80510^{\rm s}, +11^{\circ}12' 43.1310'')$ and
$ (08^{\rm h} 03^{\rm m} 51.29112^{\rm s}, +09^{\circ}57' 18.8060'')$ 
at each date, respectively.
Those positions are independent of the ephemeris, but depend on the assumed star positions
(Eqs.~\ref{eq:a_d_star_2012} and \ref{eq:a_d_star_2014}).
\end{footnotesize}
\end{deluxetable}

\clearpage

\clearpage

\clearpage

\clearpage


\clearpage
\newpage

\clearpage
\newpage

\clearpage

\clearpage

\clearpage

\end{document}